\documentclass[preprint,11pt]{elsarticle}

\usepackage{amsmath,amssymb,amsthm,amsfonts}
\usepackage{booktabs}
\usepackage{siunitx}

\usepackage[colorlinks=true,linkcolor=blue]{hyperref}
\usepackage{lineno}
\newcommand{\dstr}[2]{\frac{d {#1}} {d {#2}}}
\newcommand{\dpar}[2]{\frac{\partial {#1}} {\partial {#2}}}

\usepackage{nomencl}

\makenomenclature

\begin{document}

\begin{frontmatter}



\title{Model and method to predict the turbulent kinetic energy induced by tidal currents, application to the wave-induced turbulence}

\journal{VSI: Marine Energy Sites}

\author[m2c]{Clément Calvino}
\ead{clement.calvino@unicaen.fr}
\author[m2c]{Lucille Furgerot}
\author[lusac]{Emmanuel Poizot}
\author[lusac]{Pascal Bailly du Bois}
\author[m2c]{Anne-Claire Bennis}

\affiliation[m2c]{organization={UMR 6143 M2C, University of Caen-Normandy},
            addressline={24 rue des Tilleuls}, 
            city={Caen},
            postcode={14000}, 
            state={Normandie},
            country={France}}
\affiliation[lusac]{organization={EA 4253 LUSAC CNAM-INTECHMER, University of Caen-Normandy},
            addressline={60 rue Max-Pol Fouchet}, 
            city={Cherbourg-en-Cotentin},
            postcode={50130}, 
            state={Normandie},
            country={France}}

\begin{abstract}

%\begin{linenumbers}
A prediction model for the turbulent kinetic energy (TKE) induced by tidal-currents is proposed as a function of the barotropic velocity only, along with a robust method evaluating the different parameters involved using Acoustic Doppler Current Profiler (ADCP) measurements from Alderney Race. We find that the model is able to reproduce correctly the TKE profiles with coefficients of correlation on average higher than $0.90$ and normalised root-mean-square errors (NRMSE) less than $14\%$.
Different profiles are also tested for the mean velocity, no satisfactory prediction model is found but we are able to have decent estimates of the velocity shear and friction velocity.
Two applications are then carried out. First the turbulent budget terms are estimated and discussed. We identify the turbulent production and dissipation of TKE as the most important mechanisms, then we discuss the validity of several theoretical results derived for isotropic turbulence for this application. A strong departure for the estimation of the turbulent dissipation is notably found and explained by the turbulent anisotropy. At last the prediction model for the TKE is used to infer the wave-induced TKE. We show the importance of removing the tidal component, waves can have a strong influence down to mid-depth.
%\end{linenumbers}

\end{abstract}


\end{frontmatter}

\newpage
%----------------------------------------------------------------
\section{Introduction \label{section_intro}}
%----------------------------------------------------------------

In the context of climate change, with the necessity of reducing the carbon emissions and moving out from fossil fuel, exploitation of tidal currents with offshore tidal turbines has regained a lot of interest in the past decade. The newly proposed FloWatt (\url{https://www.flowatt.fr/}) project in France is a good example. It will allow to install a pilot tidal turbine farm in Alderney Race, the most energetic coastal site in the English Channel. The farm would be composed of $7$ turbines for a $17.5\,\si{\mega\watt}$ total installed power, with the ambition of making tidal energy a reliable and competitive source of electricity.
Accurate knowledge of the forces acting on the water turbine is crucial for predicting the generated power, dimensioning efficiently the submerged structures and ensuring a resilient integration in the continental power grid.

As a first approximation the tidal currents are the main source of energy in the water column, their interaction with the seabed generates ambient marine turbulence and shear in the column (e.g. \cite{mercier2020numerical}). 
As such it is crucial to have a good understanding of this interaction, 
%and we will remind in this introduction the basic theory on this topic.
the basic wall theory is reminded in \ref{wall_turbulence_theory}.
Realistic conditions are however often more complex than the wall theory. The conclusion from a large number of studies cited in the review paper \cite{neill2021review} indicates that the nature and intensity of turbulence is strongly dependent on the tidal conditions and local site features \cite{togneri2016micrositing}. The base hypothesis that turbulence is isotropic is also shown to be wrong in most cases, with a stronger turbulence observed in the stream-wise direction \cite{thomson2012measurements}. The wall theory is therefore hardly applicable as such, and in-situ measurements remain necessary to have a good understanding of the turbulence at one particular location.
However, other effects can have surprisingly strong impacts on the hydrodynamic variables. Waves for instance can generate additional conservative forcing on the water column through the vortex force or the Bernoulli head, or additional transport through the Stokes drift. Non-conservative effects like wave breaking have been observed to reach high depth as well and greatly impact the turbulent mixing in the upper part of the water column.
Estimating the tidal component and extracting the wave-induced component from variables of interest is therefore useful to better characterise the hydrodynamics of a renewable energy site. The purpose of the research presented here is to develop a method able to evaluate the tidal-generated TKE and velocity near the seabed, then apply the method in order to deduce the wave-induced turbulence notably.

\begin{figure}
 \centering \includegraphics[height=7cm]{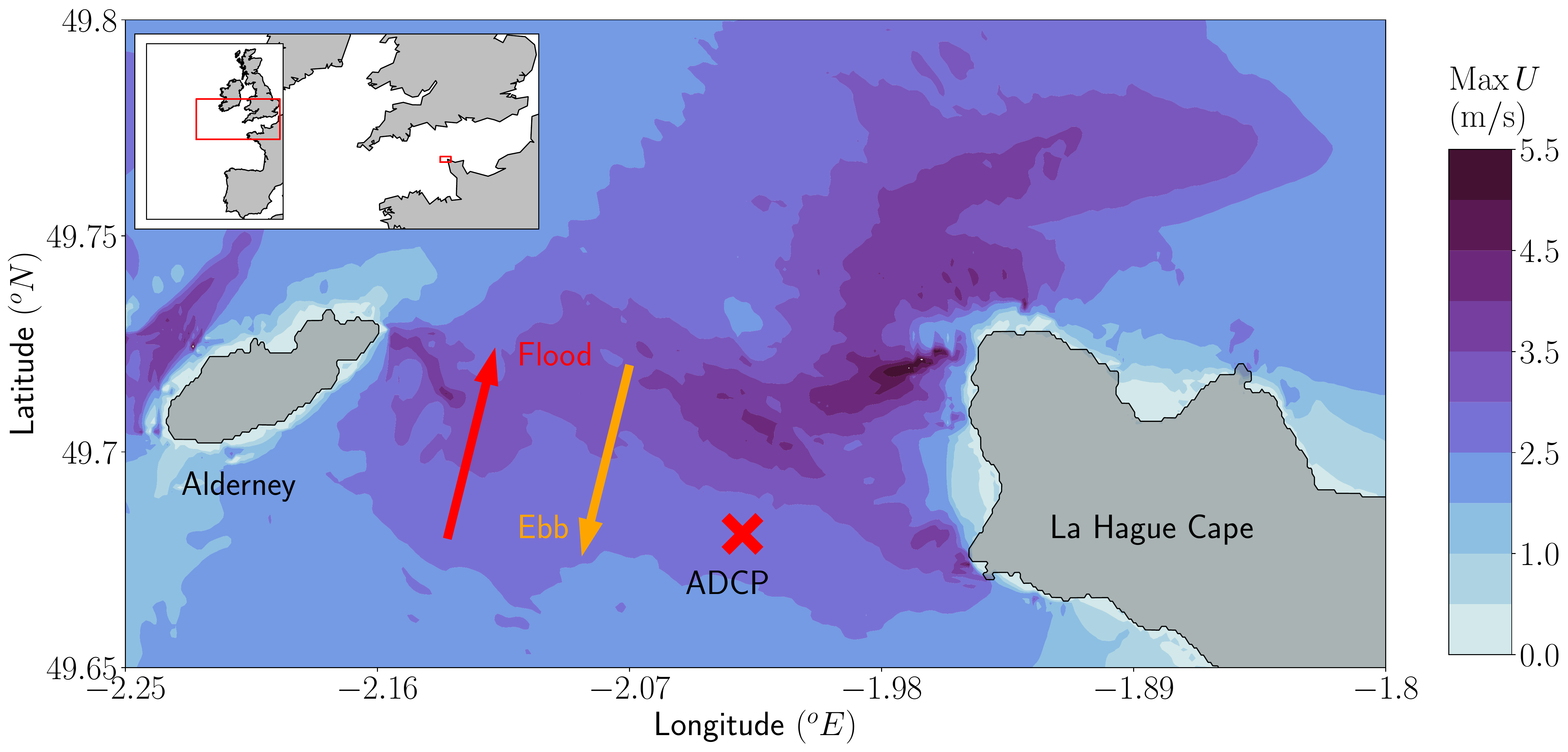}
 \caption{Map of Alderney Race showing the peak spring currents in the area using data from \cite{baillydubois2020alderney}, the calculations were performed during two moon cycles between $01/03/2015$ and $29/04/2015$; the position of the ADCP used in this research is marked by a red cross; red and orange arrows indicate the respective direction of flood and ebb flows.}
 \label{map_alderney}
\end{figure}

Tides are the main driving mechanism in Alderney Race, it is a strong macro-tidal and semi-diurnal environment \cite{baillydubois2020alderney,bennis2020numerical}. Tidal currents can reach up to $5\,\si{\m\per\s}$ as shown in Fig. \ref{map_alderney}. The interaction of those currents with the seabed generates the shear in the vertical velocity profiles, as well as marine turbulence. The most simple and yet often used theory to describe this interaction and characterise the flow is the wall turbulence theory (e.g. \cite{pope2001turbulent}).
The wall theory and consequently the logarithmic law for the velocity vertical profile have been shown to work well in the oceanic shelf boundary layer, at least for the bottom first $20\%$ of the water column \cite{perlin2005lawofthewall,lueck1997logarithmic,nezu1993turbulence}. Divergence can be observed higher in the water column as other factors influence the turbulent dynamics.

A short reminder on wall turbulence is shown in \ref{wall_turbulence_theory}.
% The previous theoretical results we have reminded all indicate
They all indicate that the velocity, the TKE but also by extension most of the budget terms appearing in Eq. (\ref{tke_balance}) can be expressed as function of the bottom friction velocity $u_\tau$ alone. The friction velocity itself can be related to the mean velocity $u$, or even the mean barotropic velocity $U$, so all the variables of interest can in theory be deduced from the tidal velocity. This feature has been observed several times in the literature. References \cite{lewis2017characteristics,thiebaut2017asymmetry} both show that a power law fitted for the velocity profile depends strongly on the tide (flood or ebb), but also on the tidal phase (increase, peak or decrease) which is directly linked to the velocity strength. Likewise Refs. \cite{togneri2017comparison,thiebaut2020assessing,furgerot2020measurements} all observe a clear correlation between the TKE and mean velocity magnitude. However less attempt have been made to characterise the vertical TKE profiles, as it is often done for the turbulent dissipation \cite{mcmillan2016rates,guerra2017turbulence,mcmillan2017spectral}.

This motivates us to present a method able to predict the velocity and turbulent kinetic energy profiles from the barotropic velocity. The end goal is to obtain a correct evaluation of the wave contribution to a variable of interest $X$, as per the rationale below:

\begin{equation}
X_{\text{wave generated}} = X_{\text{measured}} - X_{\text{tidal, predicted}} . \label{goal}
\end{equation}

The previous expectations are only relevant provided that the conditions of measurements are close to the conditions in which the theory is derived. Notably a fully developed and permanent flow is assumed, over a flat bottom with no horizontal variations. Those are strong assumptions which are not necessarily verified and will be discussed throughout the paper. Mostly inspired from the theory, the models used to predict the velocity and TKE are presented in Section \ref{section_method}. Surface effects such as wave breaking and induced mixing must be negligible and not impact the flow in a first step where we fit the models, which is a strong limitation.
%on the data usable.
We test our models and method with adequate ADCP data described in Section \ref{section_data}, only keeping measurements obtained during calm sea states and wind conditions. A first analysis of the velocity laws is presented in Section \ref{section_velocity}, in order to justify the analytical model retained. The accuracy of the two models predicting the velocity and TKE profiles is then studied in Section \ref{section_profiles}, along with a brief discussion on the limitations of such models. We finally produce two applications in Section \ref{section_applications}, notably using the fitted model to estimate the different terms in the TKE balance equation and estimate the impact of waves on the velocity and TKE profiles. We use the occasion to discuss the observed departure from the wall theory expectations.

%----------------------------------------------------------------
\section{Theoretical Modelling \label{section_method}}
%----------------------------------------------------------------

The method used to assess the ocean wave contribution by filtering out tidal effects is described here. Based on the theory exposed in introduction, we predict the depth-dependent velocity and TKE from the mean barotropic velocity only.
Figure \ref{vel_tke_error}A justifies such a choice as there is a clear one-to-one and onto relationship between the barotropic and depth-dependent velocities.
Several other choices are made and explained in this section, for compactness purposes some are justified by referencing later observations based on the data analysed, either in Section \ref{section_data}, \ref{section_profiles} or \ref{section_applications}.

\subsection{Depth-dependent velocity}

Several theoretical formulations for the velocity vertical profiles can be found in the literature. As exposed in \ref{wall_turbulence_theory} the logarithmic profile given by Eq. (\ref{log}) is the most classic expression expected. The unknowns are the friction velocity $u_\tau$ and the bottom roughness $z_0$, which are obtained for each profile through a brute optimisation algorithm, minimising the square root of the sum of squares.
A wake correction is often introduced further away from the wall, providing additional degrees of freedom to the model approximation. Such law of the wake is mostly relevant for channel or pipe flows and lacks physical interpretation \cite{krug2017revisiting}, it is therefore not preferred for free surface oceanic flows.
We will introduce two other models, the power law and the double logarithmic law, then test and discuss their accuracy to reproduce the velocity profiles later in Section \ref{section_profiles}.

Power laws are widely used by oceanographers, mostly for its convenience and efficiency in tidal energy research \cite{lewis2017characteristics}. Such a law is reminded below with Eq. (\ref{pwr}), where the unknowns optimised for each fit are the power $\gamma$ and the roughness coefficient $\delta$, obtained through a brute optimisation algorithm:

\begin{equation}
u(z) = \left(\frac{z}{\delta h} \right)^{1/\gamma} U. \label{pwr}
\end{equation}

The double logarithmic law has been introduced in \cite{sanford1999turbulent,trowbridge1999nearbottom} notably, described by Eq. (\ref{loglog}) below:

\begin{equation}
u(z) = \left\{
    \begin{aligned}
        \frac{u_{\tau,\text{bot}}}{\kappa} \ln\left(\frac{z}{z_{0,\text{bot}}}\right), && \text{if}\ z \leq z_\text{lim}, \\
        \frac{u_{\tau,\text{up}}}{\kappa} \ln\left(\frac{z}{z_{0,\text{up}}}\right), && \text{if}\ z > z_\text{lim}.
    \end{aligned}
\right. \label{loglog}
\end{equation}

The two logarithmic layers are separated by the variable $z_\text{lim}$, each layer features its own friction velocity and bottom roughness. All unknowns are determined by running a brute optimisation algorithm on all five parameters.
%, including $z_\text{lim}$.
The division in two log layers has been sometimes attributed to the action of form drag on the upper layer \cite{sanford1999turbulent}, but some have argued that it could be the result of stratification, acting as a new constraint for the eddy size \cite{trowbridge2018boundary,perlin2005lawofthewall,deserio2014streamwise}. In Ref. \cite{perlin2005lawofthewall} for instance they introduce the Ozmidov scale and use it to propose a new model for the turbulent length scale. It is however impossible to test their model here as we lack measurements of the water pressure, necessary to evaluate the Ozmidov scale. Furthermore their formulation is likely to be irrelevant in our case as the water column is barely stratified, due to the intense turbulent mixing present in Alderney Race.

The three laws mentioned above are tested in Section \ref{section_profiles} and their accuracy and performance discussed. Arguments will be given in favour of using the logarithmic law, further analysis is then carried out.
The friction velocity in the bottom boundary layer is supposed to be related to viscous processes at the bottom \cite{sanford1999turbulent}, depending solely on the flow speed \cite{pope2001turbulent}. As such a linear regression is suited to scale this variable as follows:

\begin{equation}
u_{\tau} = a_{\tau} U + b_{\tau} .
\end{equation}

The roughness heights are attributed to characteristics of the seabed for high Reynolds number, as reviewed in \cite{trowbridge2018boundary} referencing \cite{jimenez2004turbulent} and \cite{allen2007turbulent}. They are not supposed to depend on the flow speed. Using the previous linear expression the following model for the depth-dependent velocity will be retained:

\begin{equation}
u = \frac{\left(a_{\tau} U + b_{\tau}\right)}{\kappa} \ln{\left(\frac{z}{z_0}\right)} . \label{v_model}
\end{equation}

The vertical velocity profile has been less documented. It is usually neglected compared to the horizontal velocities, but it can still impact the vertical advection of TKE. The logarithmic law introduced in Section \ref{section_intro} assumes a horizontally homogeneous flow. Combined with the continuity equation it yields a constant vertical velocity, which is too restrictive and is not agreeing with observations.
Still we decided not to model the vertical velocity by lack of theoretical background providing an analytical form for the vertical velocity, and by lack of evident scaling with respect to the barotropic velocity appearing on the profiles (e.g. \cite{mercier2020numerical}).

\subsection{Depth-dependent turbulent kinetic energy}

The classic wall theory and similar works suggest that the TKE is first constant within the logarithmic layer, which spawns the first few meters of the water column, and then decreases as a linear function of the height above seabed. It is traditionally assumed to scale as $u_\tau^2$, or equivalently as $U^2$, but we wish to challenge this dimensional argument.

For each profile the linear regression given by Eq. (\ref{reg_1}) is computed for $k_t$:

\begin{equation}
 k_t = \alpha z + \beta . \label{reg_1}
\end{equation}

This first regression gives us samples for $\alpha$ and $\beta$, which depend primarily on the barotropic velocity. Regressions for the coefficients $\alpha$ and $\beta$ are then conducted to fit a power model given by Eq. (\ref{reg_2}) and suggested by the theory (e.g. \cite{pope2001turbulent}):

\begin{equation}
\alpha = a_\alpha {U}^{p_{\alpha}} + b_\alpha \, \qquad \text{and} \qquad \beta = a_\beta {U}^{p_{\beta}} + b_\beta . \label{reg_2}
\end{equation}

The regressions are done separately, if the TKE indeed scales with the mean flow then similar values for $p_\alpha$ and $p_\beta$ should be found. It will be verified in Section \ref{section_applications}. A mean value $p$ is kept afterwards, which still depends on the tidal phase:

\begin{equation}
 p = (p_\alpha + p_\beta)/2 . \label{reg_3}
\end{equation}

The goal is now to obtain reference profiles for the TKE, given the scaling found previously. It can be expressed as follows, with $A(z)$ a scaling function varying with depth and $k_{t,0}(z)$ a residual TKE:

\begin{equation}
 k_t(z) = A(z) {U}^{p} + k_{t,0}(z) . \label{kt_model}
\end{equation}

At this stage it is already possible to find expression for $A(z)$ and $k_{t,0}(z)$ using Eqs. (\ref{reg_1}--\ref{reg_2}):

\begin{equation}
 A(z) = a_\alpha z + a_\beta , \qquad k_{t,0}(z)  = b_\alpha z + b_\beta . \label{reg_4}
\end{equation}

The solution proposed with Eq. (\ref{reg_4}) forces a linear profile for the TKE, it is certainly a decent approximation close to what the theory suggests. However such an approach is still forcing theoretical results on a complex application where several hypotheses are not verified, such as horizontal homogeneity. Consequently linear regressions are then conducted bin per bin, at each height $z$. The TKE is estimated with regards to the scaled tidal velocity ${U}^p$. Those regressions directly yield the slope $A(z)$ and the intercept $k_{t,0}(z)$, with no underlying hypothesis on their vertical shape.

Once the three parameters $A(z)$, $k_{t,0}(z)$ and $p$ are evaluated, it is possible to predict the TKE generated by the tidal currents as long as the barotropic velocity is known. The accuracy of the method is evaluated in Section \ref{section_applications}.

%----------------------------------------------------------------
\section{Data \label{section_data}}
%----------------------------------------------------------------

\begin{table}
\begin{center}
\begin{tabular}{l  c}
\toprule
 \textbf{Instrument}     & \textbf{ADCP Sentinel V50 5 beams} \\
 \textbf{Campaign}      & \textbf{Leg56} \\
\midrule
 Deployment period & $27/02/2018$--$06/07/2018$   \\
 Latitude $(\si{\degree} N)$  & $49.68100$ \\
 Longitude $(\si{\degree} E)$ & $-2.02965$ \\
 Water depth (MSL) $(\si{\m})$ & $37.7$ \\
 Beam frequency $(\si{\kilo\Hz})$ & $500$\\
 Sampling frequency $(\si{\Hz})$ & $2$ \\
 Burst duration $(\si{\min})$  & $20$ \\
 Burst frequency $(\si{\min})$ & $60$ \\
 Vertical resolution $(\si{\m})$ & $1$ \\
 Range $(\si{\m})$ & $2.7$--$27.7$ \\
 Beam inclination $(\si{\degree})$ & $25$ \\
\midrule
 \textbf{Number of samples} &  \\
\midrule
 Flood cases                                & $1536$ \\
 \qquad with $H_s < 0.7\,\si{\m}$           & $642$  \\
 \qquad with $H_s < 0.7\,\si{\m}$ and $U > 1.5\,\si{\m\per\s}$       & $289$  \\
 Ebb cases                                  & $1456$ \\
 \qquad with $H_s < 0.7\,\si{\m}$           & $567$  \\
 \qquad with $H_s < 0.7\,\si{\m}$ and $U > 1.5\,\si{\m\per\s}$       & $235$  \\
\bottomrule
\end{tabular}
\end{center}
\caption{ADCP set-up parameters, as well as a breakdown of $20\,\si{\min}$ long samples available.}\label{adcp_param}
\end{table}

The data used in this paper are extracted from a five-beam bottom-deployed ADCP in Alderney Race between $27/02/2018$--$06/07/2018$, a map of the area with the position of the ADCP is shown in Fig. \ref{map_alderney} with a summary of its configuration given in Table \ref{adcp_param}. More details on the configuration is given in \cite{furgerot2020measurements}, where it is referred as ADCP S1 Leg56, they also include in their paper more ADCP and high-frequency radar measurements not exploited in the present research.

The raw high-resolution beam velocities and sea level are available, recorded with a $2\,\si{\hertz}$ frequency during $20\,\si{\min}$ burst windows. Usable bin heights are between $2.7\,\si{\m}$ and $17.7\,\si{\m}$ or $27.7\,\si{\m}$ above the sea bottom, depending on the velocity speed. With higher velocities the quality of measurement improves and more bin heights are available. Outside of those ranges data show too much contamination or too low correlation and amplitude intensity. The bin resolution is $1\,\si{\m}$.
The measured mean total depth is $37.7\,\si{\m}$ with a maximum tidal range of $8.2\,\si{\m}$ during the duration of measurements, which includes $9$ spring and neap tides.

A quality analysis is first conducted to remove unrealistic records, sources of errors include lobe contamination, free surface contamination and oscillations of the structure above the recommended range. The data is notably despiked following \cite{goring2002despiking}, although only one sweep is carried out to remove spikes for running time purposes.
The high-resolution data is then projected in East-North-Up directions and averaged over the $20\,\si{\min}$ burst window length to obtain mean statistics. The calculation of the turbulent statistics is detailed below.

The height above seabed is the preferred vertical coordinate here as we are mostly interested in the sea bottom effects and generation of turbulence. This coordinate is constant for each ADCP bin. However when studying the wave-induced currents and TKE we prefer to use the relative depth, computed as the mean sea level minus the height above seabed. Such coordinate varies for a given ADCP bin.

We briefly describe how the raw high-frequency data are processed for each of the variables of interest. We also explain some of the choices made with regard to which variable is used for the analysis.

\subsection{Sea level}

The raw data for the sea level consist of a time series obtained with the surface tracking method. The sea level is notably used to compute the position of the bin below the surface and in the wave orbital spectra calculation, through the wave dispersion relation. Since the water depth is large compared to the tidal range we are not interested in the fast oscillations induced by the waves. The sea level is then simply averaged over the $20\,\si{\min}$ burst record window. 
As a result the mean sea level is still capturing any adjustment to local atmospheric or wave conditions. Given the precision of the acquisition ($0.33\,\si{\m}$), the beam accuracy ($1\,\si{\m}$) and the average depth of the first usable bin below the sea surface ($6\,\si{m}$), we do not think that those local surface forcing processes impact the sea level drastically, nor the analysis.

\subsection{Velocities}

\begin{figure}
 \centering \includegraphics[height=7cm]{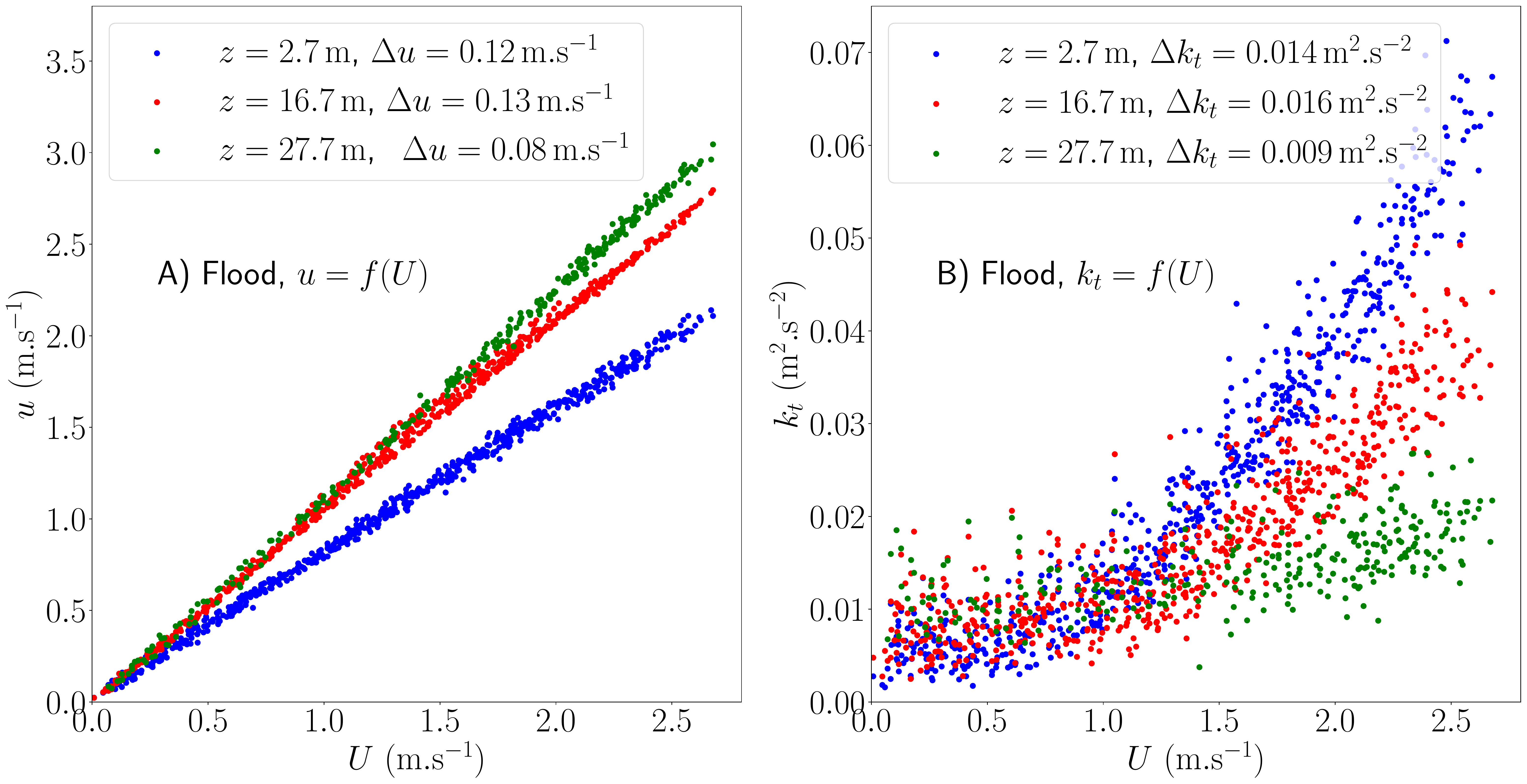}
 \caption{Plot A shows scatter graphs of the mean depth-dependent horizontal velocity with respect to the barotropic horizontal velocity for flood cases at three different heights above seabed ($27.7\,\si{\m}$, $13.7\,\si{\m}$ and $2.7\,\si{\m}$), an estimation of the uncertainty is given with $\Delta u$ evaluated as the $95\%$ prediction interval for each corresponding linear regression; plot B shows the TKE with respect to the barotropic velocity at the same depths, uncertainty is estimated with $95\%$ prediction intervals as well modelling the TKE with a squared power law ($k_t = A U^2 + B$).}
 \label{vel_tke_error}
\end{figure}

The mean velocity is obtained by conducting averages over the $20\,\si{\min}$ record windows. We still suspect turbulent oscillations to be present in the data after average, as illustrated with Fig. \ref{vel_tke_error}A. The estimation of the error is carried out by computing the $95\%$ prediction intervals after performing a linear regression of the mean three-dimensional velocity with respect to the mean barotropic velocity, yielding at most $0.13\,\si{\m\per\s}$. It is similar to the value found in \cite{sanford1999turbulent} with their own instrument, recording at a $1\,\si{\hertz}$ frequency.
The dispersion of the data is not necessarily caused by remnant turbulence only, other unknown processes could have an influence such as the tidal phase or tidal coefficient. We still believe that turbulence is the major cause.

The barotropic velocities are computed by depth-averaging each velocity component over the vertical range measured by the instrument. Given the mean total depth ($37.7\,\si{\m}$) only $68\%$ of the water column roughly is used for the depth-average.
%The quantities obtained are then not exactly the barotropic velocities, but we still refer to them as such and this approximation is not damaging the analysis.

\subsection{Turbulent kinetic energy}

The TKE is computed following the formula given in \cite{dewey2007reynolds} and reminded below, with $b'_i$ the turbulent velocity along beam $i$, with the fifth beam being the vertical beam, $\theta$ is the beam inclination and $\phi$ the pitch angle in radians:

\begin{equation}
k_t = \frac{1}{4 \sin^2{\theta}} \left( \overline{b_1^{'2}} + \overline{b_2^{'2}} + \overline{b_3^{'2}} + \overline{b_4^{'2}} - 2(2 \cos^2{\theta} - \sin^2{\theta}) \overline{b_5^{'2}} - (\cot{\theta} -1)\phi(\overline{b_2^{'2}} - \overline{b_1^{'2}}) \right). \label{tke_dewey}
\end{equation}

The mean beam velocity is obtained with an average over the $20\,\si{\min}$ long record windows, the turbulent component is then computed by removing this mean value from the instantaneous beam velocity. By definition the TKE is then a mean statistics over the $20\,\si{\min}$ of each record sample. Similarly to what is observed for the mean velocity some dispersion still appears in the data, attributed to the turbulent unsteadiness of the flow. It is illustrated with Fig. \ref{vel_tke_error}B only showing calm sea states during flood cases. This inherent error carried in the data is estimated around $0.01\,\si{\square\m\per\square\s}$, after computing the $95\%$ prediction intervals using a simple squared law for the TKE ($k_t = A U^2 + B$).

Doppler noise filtering can be carried out to remove measurement noise, for instance it is done in \cite{mcmillan2016rates,thiebaut2020comprehensive} following the noise auto-correlation approach exposed in \cite{durgesh2014noise}. However we decided not to conduct it here. This technique requires to identify the inertial subrange on the velocity spectra, expected to exhibit a $-5/3$ slope, as well as a flattening of the spectra after this inertial subrange. Over only one $20\,\si{\min}$ burst window the computed spectra are not smooth enough to identify clearly those two regions, we therefore decided not to apply the correction. Furthermore the method presented in Section \ref{section_method} is estimating an error parameter $k_{t,0}(z)$ for each profile, which by construction includes the Doppler noise contribution.

\subsection{Turbulent dissipation \label{integral_method}}

In the one-equation TKE balance given by Eq. (\ref{tke_balance}), the dissipation $\epsilon$ is evaluated from $k_t$ itself and the mixing length (Eq. \ref{epsilon_kt}). The dissipation can however be estimated independently from the velocity spectral density, as done for instance in \cite{guerra2017turbulence,mcmillan2016rates}. Several methods are also available to remove the wave orbital contribution from the velocity spectra, well described in \cite{thiebaut2020comprehensive,filipot2015wave}. As recommended we decided to opt for the integral method. In the isotropic inertial subrange the turbulent velocity spectral density $S(k)$ follows the decay law given below, with $k$ the wavenumber of the turbulent structures in the flow, $C$ a vertical Kolmogorov constant, $N$ the instrument Doppler noise and $S_w$ the wave orbital spectral density:

\begin{equation}
 S(k) = C \epsilon^{2/3} k^{-5/3} + N(k) + S_w(k). \label{spectrum_epsilon}
\end{equation}

With the data available two independent estimation of $\epsilon$ are possible, either with the vertical ADCP beam or with the four inclined beams. The choices for $S$, $N$ and $C$ are reminded with Table \ref{dissip}, as mentioned in \cite{mcmillan2016rates}.

For each $20\,\si{\min}$ record sample the beam velocity spectra are computed by averaging $3\,\si{\min}$ long segments overlapping at half their length. The correction with the wave orbital spectra is then carried out. They are estimated from the wave spectra using the linear theory, with adequate projections in order to match the beam velocity directions. The equations are fully described in \ref{wave_orbital_app}.
In order to smooth further the wave corrected spectra they are also averaged by velocity bins of width $0.4\,\si{\m\per\s}$.
The Doppler noise is then evaluated by averaging the tail of the spectra at each depth and each velocity bin, following \cite{thiebaut2020comprehensive}, where the instrument noise is found to increase with the tidal current velocity. In \cite{mcmillan2016rates} a unique value is used for all velocity values, as a unique Doppler floor value is observed at high frequency regardless of the velocity. We suspect however that the higher frequency sampling of their vertical beam ($8\,\si{\hertz}$) makes this observation available, but for a lower frequency sampling ($2\,\si{\hertz}$) the noise is affected by the tidal velocity.

Once all the corrections are done, Eq. (\ref{spectrum_epsilon}) is used to infer the dissipation $\epsilon$. The corrected spectra are transformed in wavenumber spectra using Taylor's frozen hypothesis. Such assumption is in general valid for strong $u$ flows, but might be more hazardous during slack phases or even near the bottom floor where friction reduces considerably the flow. We still conduct it for all available data.
The compensated wavenumber spectra are then computed multiplying by $k^{5/3}$, and the local maximum is identified on each spectrum in order to estimate $\epsilon$.

\begin{table}
\begin{center}
\renewcommand{\arraystretch}{2}
\begin{tabular}{l  c  c}
\toprule
 Approach & Four beam & Vertical beam \\
\midrule
 $C$ & $2 C_u \sin^2 {\theta} + 2 C_w \sin^2 {\theta} + 4 C_w \cos^2 {\theta} $ & $C_w$ \\
 $S(k)\,(\si{\square\m\per\square\s}/\si{\radian\per\m})$ & $\displaystyle \left[S_{11}(f) + S_{22}(f) + S_{33}(f) + S_{44}(f)\right] \frac{u}{2\pi}$ & $\displaystyle S_{55}(f) \frac{u}{2\pi}$ \\
 $N(k)\,(\si{\square\m\per\square\s}/\si{\radian\per\m})$ & $\displaystyle \left[N_{1}(f) + N_{2}(f) + N_{3}(f) + N_{4}(f)\right] \frac{u}{2\pi}$  & $\displaystyle N_{5}(f) \frac{u}{2\pi}$ \\
\bottomrule
\end{tabular}
\end{center}
\caption{Breakdown of the terms in Eq. (\ref{spectrum_epsilon}), the spectra from the beam velocities ($S_{ii}(f)$) are converted from frequency to wavenumber spectra using the frozen field approximation, $C_u=0.49$ and $C_w=0.69$ are the streamwise and transverse Kolmogorov constants.}\label{dissip}
\end{table}

\subsection{Wave parameters}

Two-dimensional frequency and directional wave spectra are extracted from the ADCP data for each record sample as well using the wave orbital technique. This step is carried out using the Teledyne Velocity software \cite{terray1999measuring}, no further detail will be given here.

\subsection{Data sampling}

The velocity profiles and current-induced TKE are evaluated for flood and ebb separately. This is necessary as the velocities and generally the behavior of all variables vary too much in each case \cite{furgerot2020measurements,togneri2017comparison}, probably due to the different bathymetric features and the history of current prior to the ADCP location in each case \cite{furgerot2019highresolution}.

A breakdown of the samples available is given in Table \ref{adcp_param}.
In order to quantify and later predict the base TKE only attributed to tidal currents only calm sea states are kept, characterised by the constraint $H_s < 0.7 \,\si{\m}$. This value is deemed the best compromise between having enough samples to conduct meaningful averages, and minimising the impact of waves on the data.
In those cases wind velocities are always less than $10\,\si{\m\per\s}$ and on average around $4\,\si{\m\per\s}$. As such any impact of the wind or waves on the velocity and TKE profiles can be neglected, such as the Ekman current, the Stokes drift, the vortex force or Stokes-Coriolis force.
As it will be argued later, cases with low velocities tend to show more variability and depart from the theoretical framework in which the models are derived. As such the statistics are less exploitable for low velocities, we decide then to recompute them with a constraint on the velocity and remove cases with $U<1.5\,\si{\m\per\s}$. The limit is arbitrary and found after several trials as a best compromise, the number of remaining samples is shown in Table \ref{adcp_param} as well.

The definition of the statistics used in this section is reminded in \ref{statistics_def}. They are computed per profile, and then averaged over all profiles.

%----------------------------------------------------------------
\section{Comparing the velocity laws \label{section_velocity}}

\subsection{Mean statistics}

The logarithmic law (Eq. \ref{log}), power law (Eq. \ref{pwr}) and double logarithmic law (Eq. \ref{loglog}) are compared with the aim of deciding on the best model capturing the vertical velocity profiles.
%The logarithmic law (Eq. \ref{log}), power law (Eq. \ref{pwr}) and double logarithmic law (Eq. \ref{loglog}) are now compared.
The statistics averaged over depth and over all profiles during calm conditions with $U>1.5\,\si{\m\per\s}$ are shown in Table \ref{vel_stats}.
%There is a clear indication on both the root-mean-square errors (RMSE) and correlation coefficients ($R$) that the double logarithmic law captures better the velocity profiles, although all three models give excellent statistics.
All three models give excellent statistics, although both the root-mean-square errors (RMSE) and correlation coefficients ($R$) are better for the double logarithmic law.
It is also worth pointing out the statistics for the logarithmic and power laws are quite similar. A better fit is observed with the power law for flood cases but it is worse for ebb cases. The logarithmic law gives more homogeneous statistics between flood and ebb cases.

\begin{table}
\begin{center}
\begin{tabular}{l  c  c  c  c  c  c  c  c  c}
\toprule
Statistics    & \multicolumn{2}{l}{Log law} & \multicolumn{2}{l}{Power law} &  \multicolumn{2}{l}{Double log law} \\
              & RMSE & $R$ & RMSE & $R$ & RMSE & $R$ \\
              & ($\si{\m\per\s}$) & & ($\si{\m\per\s}$) & & ($\si{\m\per\s}$) & \\
\midrule
Flood        & $0.019$ & $0.959$ & $0.011$ & $0.958$ & $0.004$ & $0.957$ \\
Ebb          & $0.015$ & $0.971$ & $0.022$ & $0.964$ & $0.005$ & $0.978$ \\
\bottomrule
\end{tabular}
\end{center}
\caption{RMSE and correlation coefficient $R$, averaged over depth, for the logarithmic law (Eq. \ref{log}), power law (Eq. \ref{pwr}) and double logarithmic law (Eq. \ref{loglog}); the statistics are computed only for calm sea states with $H_s < 0.7 \,\si{\m}$ and $U>1.5\,\si{\m\per\s}$; the bias is not shown as it is near $0$ for all cases.} \label{vel_stats}
\end{table}

\subsection{Analysis of specific profiles}

\begin{figure}
 \centering \includegraphics[height=14cm]{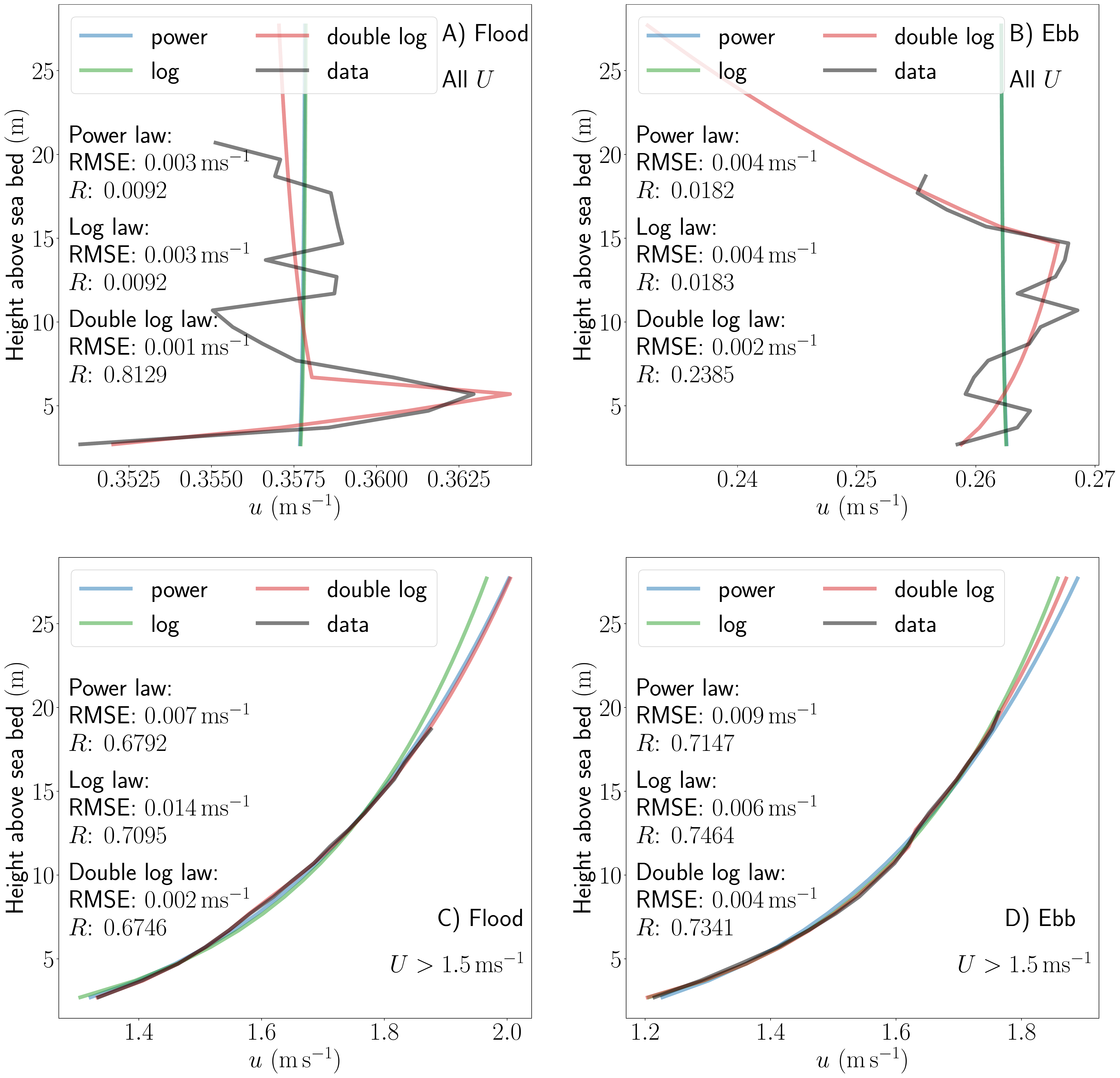}
 \caption{Horizontal mean velocity profiles (black lines), with their logarithmic law (green line), power law (blue line) and double logarithmic law profiles (red line); plots (A, B) correspond to the worst agreement found in all the available data, for plots (C, D) the constraint $U>1.5\,\si{\m\per\s}$ is added; flood cases are shown with plots (A, C) while ebb cases with plots (B, D); on all plots the log and power laws are almost overlapping identically, the profiles are extrapolated to the surface for readability only but only relevant up to the last measured data point}.
 \label{vel_profiles}
\end{figure}

For further investigation and understanding of the statistics four profiles are exhibited in Fig. \ref{vel_profiles} when the logarithmic law shows the worst agreement in terms of correlation coefficient.

For the upper row all profiles are included and the worst agreement is found near slack cases both during flood and ebb flows. The RMSE are still small as the amplitudes are small as well, with variations of the order of $0.01\,\si{\m\per\s}$. On the other hand the correlation coefficients are bad as the shapes of the profiles are not well captured, especially by the simple logarithmic and power laws. It is not surprising as during the slack period the flow turns and is not well developed. The conditions are therefore far from the hypotheses assumed in the theory and there is no reason for the profiles to agree with either a logarithmic or power vertical profile. The difference in statistics between the double logarithmic model and the two others is significant with RMSE found $3$ to $4$ times lower for the former. However this seemingly good agreement mostly comes from the piece-wise definition of the double logarithmic law, allowing for more degrees of freedom and making it easier to fit the strong discontinuity in shear observed in the data.

The lower row in Fig. \ref{vel_profiles} only keeps cases with $U>1.5\,\si{\m\per\s}$, the profiles are more developed, smoother, and agree well with their theoretical expectations. The comparison between the three models is therefore meaningful, and it justifies the choice made in Table \ref{vel_stats} to keep such cases as well. The double logarithmic law still gives significantly better agreements as it can adjust better to a change in velocity shear whereas the simple logarithmic law or power law are only able to give an average estimation of the shear stress. This is usually interpreted as a change in dominant mechanism generating the turbulence \cite{perlin2005lawofthewall,deserio2014streamwise}.
A clear decrease in velocity shear can be observed around $13\,\si{\m}$ on both Fig. \ref{vel_profiles}C and \ref{vel_profiles}D. However it appears as a discontinuity in shear on the double logarithmic law profile, which is not a realistic behavior.

\subsection{Dependence of the regression parameters}

Although the two other laws are unable to fit correctly both the lower and upper parts of the velocity profile we still decide to discard the double logarithmic law, in order to keep smooth profiles. The coefficients of the regressions are now compared to study their dependence towards the barotropic velocity $U$.

\begin{figure}
 \centering \includegraphics[height=7cm]{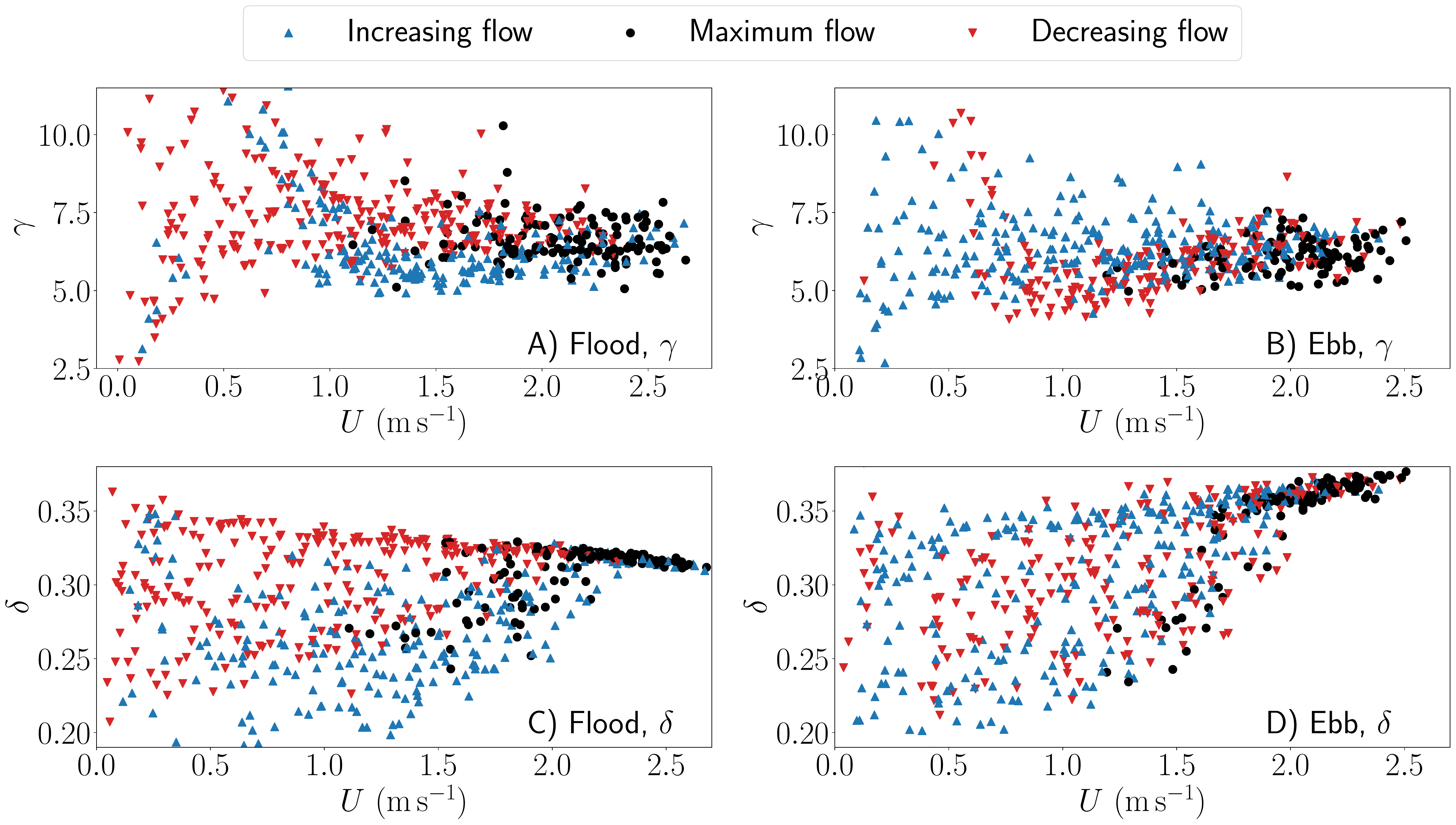}
 \caption{Scatter plots of $\gamma$ (plots A and B) and $\delta$ (plots C and D) parameters appearing in Eq. (\ref{pwr}) with respect to the barotropic velocity, for both flood (plots A and C) and ebb (plots B and D) cases; colours indicate the timing of the tide, blue corresponds to the beginning of the tide where the flow is increasing, black to the maximum of the tidal flow and red to the end of the tide with a decreasing flow.}
 \label{pwr_coef}
\end{figure}

We start with the power law coefficients plotted in Fig. \ref{pwr_coef}. The values are in the same range than found in \cite{lewis2017characteristics}, although we obtain slightly lesser bed roughness coefficients.
The exponent $\gamma$ depends weakly on $U$ overall, coefficients are estimated between $5$ and $7.5$ overall with a strong influence and spread for low velocities. However the timing of the tide impacts the power law estimation, especially for $U>1.5\,\si{\m\per\s}$. At the beginning of the tidal flood lower values of $\gamma$ are found, between $5$ and $7$, compared to the end of the tide, between $6$ to $9$. The opposite trend is observed for the ebb cases, although with less discrepancy overall.
The roughness coefficient $\delta$ shows a small trend with respect to $U$ with still a strong spread for $U < 2\,\si{\m\per\s}$. During flood cases $\delta$ decreases as the flow becomes stronger, from $0.33$ to $0.31$, while it increases during ebb cases, from $0.35$ to $0.38$. Because of the weak overall trend of $\gamma$ and $\delta$ with respect to $U$, and since the power law bears no physical meaning, we decide to discard it for the remainder of the analysis.

\begin{figure}
 \centering \includegraphics[height=7cm]{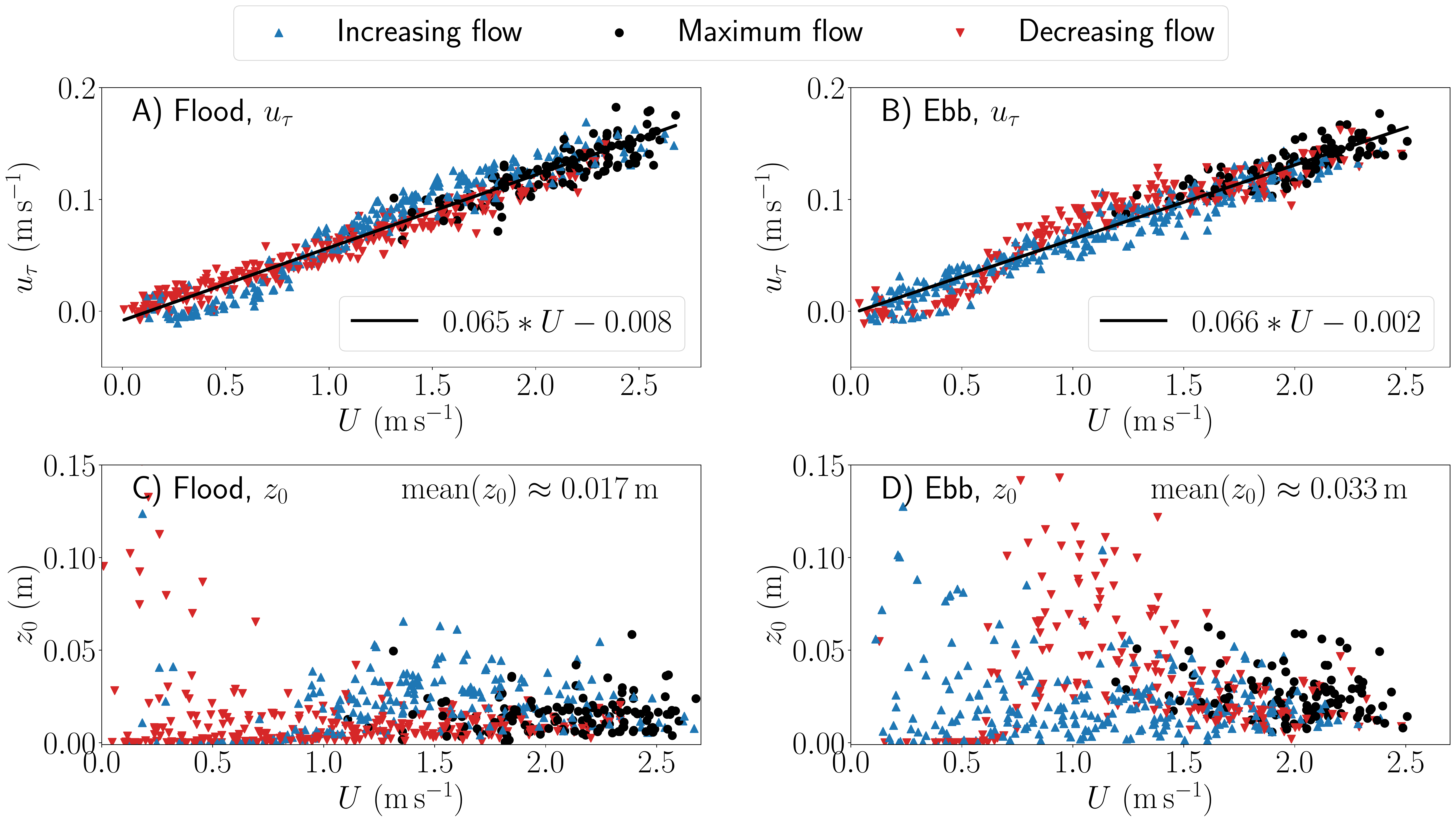}
 \caption{Scatter plots of $u_\tau$ (plots A and B) and $z_0$ (plots C and D) parameters appearing in Eq. (\ref{log}) with respect to the barotropic velocity, for both flood (plots A and C) and ebb (plots B and D) cases; colours indicate the timing of the tide, blue corresponds to the beginning of the tide where the flow is increasing, black to the maximum of the tidal flow and red to the end of the tide with a decreasing flow; the black line in plots A and B is the linear regression, with coefficients indicated in-plot.}
 \label{log_coef}
\end{figure}

The logarithmic law coefficients are now described with Fig. \ref{log_coef}.
The bottom friction velocities $u_{\tau}$ presented on plots A and B range between $0$ and $0.2\,\si{\m\per\s}$, a clear increasing trend towards $U$ is noticed with small differences due to the tidal phase. Linear regressions are computed and plotted for flood and ebb cases, the intercept is reasonably low in both cases and a similar slope is found around $0.065$. This value is coherent with previous research (e.g. \cite{stacey1999measurements}).
The roughness heights are shown with plots C and D. Overall they have a lesser dependence with respect to $U$, with the range of values varying strongly but remaining below $0.10\,\si{\m}$ in almost all cases. Similar values are found by \cite{cheng1999estimates} for instance, but contrary to our observation they obtain a clear negative correlation with respect to the mean flow.
Given the resolution and vertical accuracy of the measurements using a mean value is relevant in our case, and we can even neglect it when it is compared to the bin height since the first bin height available is $2.7\,\si{\m}$.

%----------------------------------------------------------------
\section{Predicting the velocity and TKE profiles: results and discussion \label{section_profiles}}
%----------------------------------------------------------------

\subsection{Velocity vertical profiles}

Using the regressions in Fig. \ref{log_coef} for the friction velocities and the mean values for the bottom roughness, the analytic model given by Eq. (\ref{v_model}) is fully determined and depends only on the barotropic flow $U$ and the tidal phase.
%Using the regressions shown in Fig. \ref{log_coef} for the friction velocities and the mean values for the bottom roughness, analytic models are built for the horizontal velocities, depending only on the barotropic flow $U$ and the tidal phase.
Given how close the coefficients are found we could almost use only one set of coefficients for both flood and ebb, it is not done since the authors believe that in general such a behavior is not to be expected due to differences in the flood and ebb flow conditions, such as the bathymetric features for instance.
%For completeness the full prediction model for the velocity is then given below:

%\begin{equation}
%u = \frac{\left(a_{\tau} U + b_{\tau}\right)}{\kappa} \ln{\left(\frac{z}{z_0}\right)} . \label{v_model}
%\end{equation}

The accuracy of the full prediction model is given in Table \ref{stats_vel}. 
Compared to the statistics of the initial logarithmic regressions in Table \ref{vel_stats}, a significant loss of accuracy is observed, even keeping only developed profiles with velocities $U > 1.5\,\si{\m\per\s}$. It is expected since a more general velocity model is used, with only three degrees of freedom for each tidal subset, whereas the statistics obtained with Table \ref{vel_stats} feature one fit per profile, meaning two degrees of freedom per profile and twice the number of samples degrees of freedom per tidal subset.
Since the friction velocity is very well explained by the flow, the problem lies in the approximation made for the roughness height $z_0$. The logarithmic profile is very sensitive towards this value and too much variability is observed for this parameter. As stated in \cite{bauer1992sources} the error and dispersion in estimating $z_0$ increases drastically when the ratio between $U$ and $u_\tau$ is high, which is the case here. \cite{cheng1999estimates} also observed a strong spread in their measurement of $z_0$ with a ratio $100$ between the minimum and maximum values. By lack of obvious trend we have no better option than to take the mean value for predicting the velocity profiles.
The RMSE are very close to the bias errors, meaning that bias is driving the error. It results in a shift, on average negative, compared to the measurements.

\begin{table}
\begin{center}
\begin{tabular}{l c  c  c  c | c  c  c  c}
\toprule
 & \multicolumn{3}{l}{$0\,\si{\m\per\s}<U<2.8\,\si{\m\per\s}$} & & & \multicolumn{3}{l}{$1.5\,\si{\m\per\s}<U<2.8\,\si{\m\per\s}$} \\
 & Bias & RMSE & $R$ & & & Bias & RMSE & $R$              \\
 & ($\si{\m\per\s}$) & ($\si{\m\per\s}$) & & & & ($\si{\m\per\s}$) & ($\si{\m\per\s}$) &   \\
\midrule
Flood        & $-0.058$ & $0.065$ & $0.83$ & & & $-0.036$ & $0.046$ & $0.96$ \\
Ebb          & $-0.10$ & $0.10$ & $0.84$ & & & $-0.12$ & $0.12$ & $0.97$ \\
\bottomrule
\end{tabular}
\end{center}
\caption{Bias, RMSE and correlation coefficient $R$ for the velocity model defined with Eq. (\ref{v_model}); averaged over all profiles on the left part of the table while only over cases with $U>1.5\,\si{\m\per\s}$ on the right part.}\label{stats_vel}
\end{table}

\subsection{Turbulent kinetic energy vertical profiles}

\begin{figure}
 \centering \includegraphics[height=7cm]{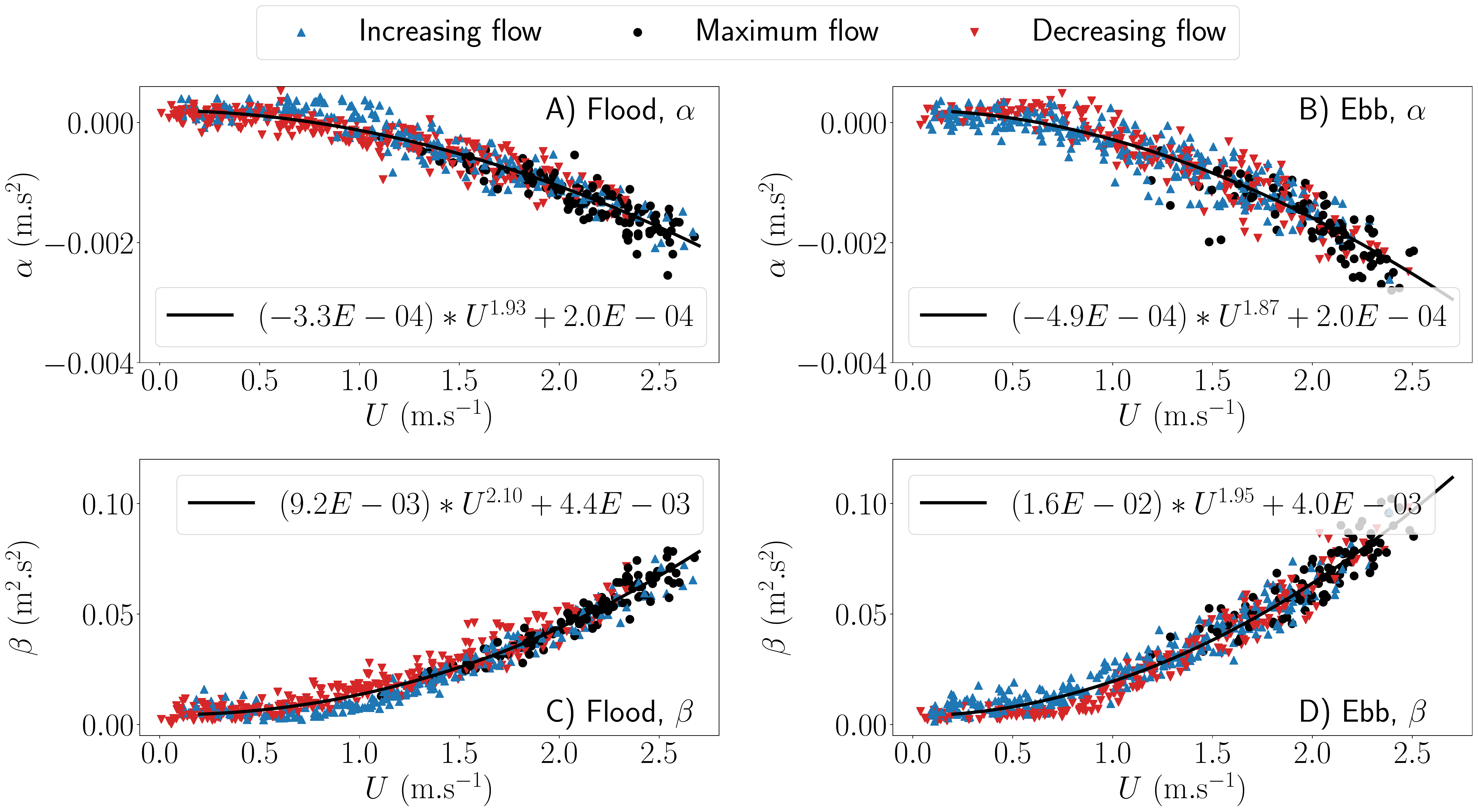}
 \caption{Coefficient $\alpha$ (plots A and B) and $\beta$ (plots C and D) obtained from the regression $k_t(z) = \alpha z + \beta$ as a function of the horizontal barotropic velocity, the flood cases (plots A and C) and ebb cases (plots B and D); power fits given by Eq. (\ref{reg_1}) are shown by the solid curves}
 \label{alpha_beta}
\end{figure}

\begin{figure}
 \centering \includegraphics[height=7cm]{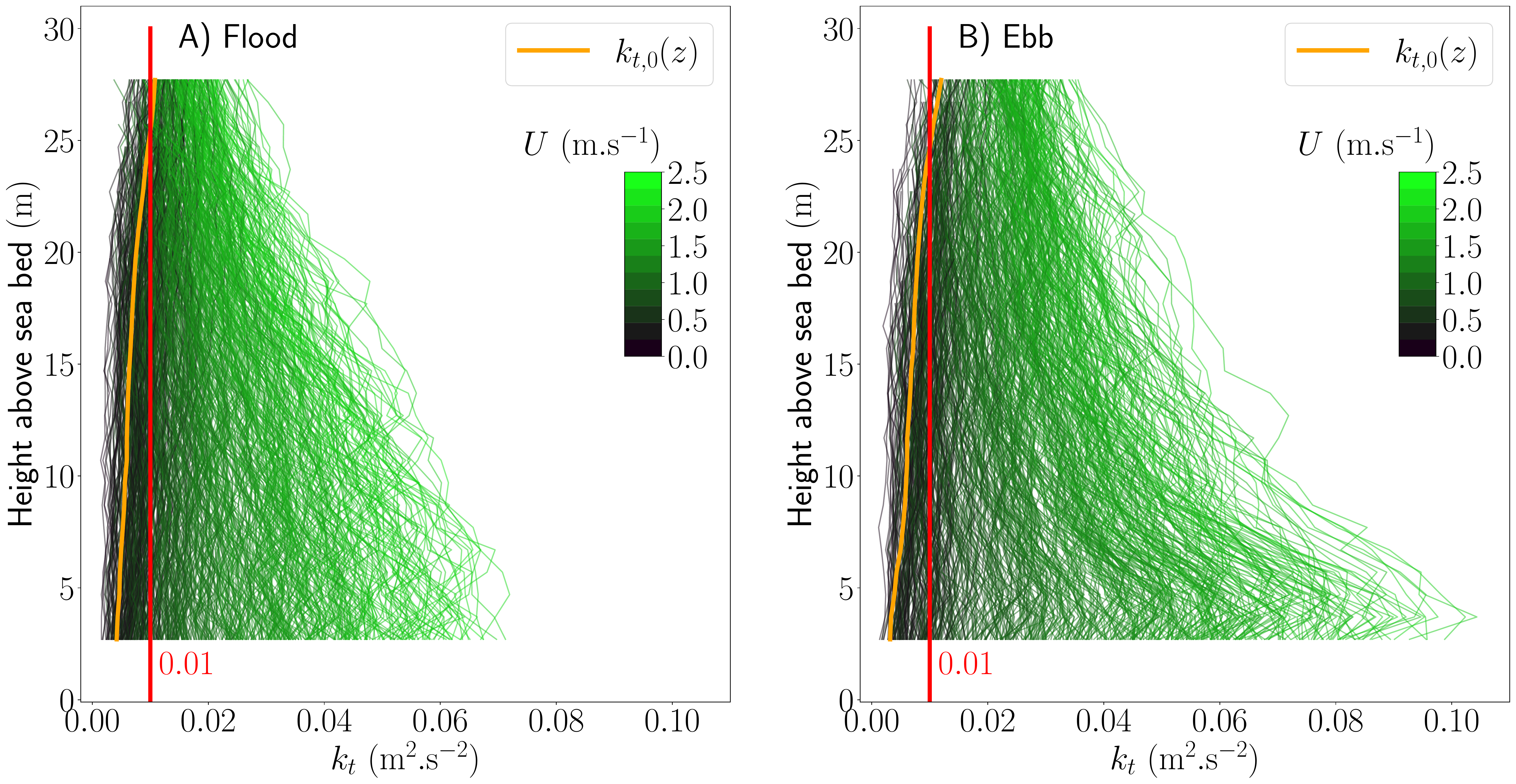}
 \caption{TKE profiles $k_t(z)$ for flood (plot A) and ebb (plot B) cases coloured with respect to the barotropic velocity, the orange profile corresponds to the intercept $k_{t,0}(z)$, interpreted as some residual TKE error; the red profile is a constant at $0.01\,\si{\square\m\per\square\s}$ and corresponds to an estimate of the error attributed to remnant unsteadiness still present in the TKE data}
 \label{tke}
\end{figure}

\begin{figure}
 \centering \includegraphics[height=7cm]{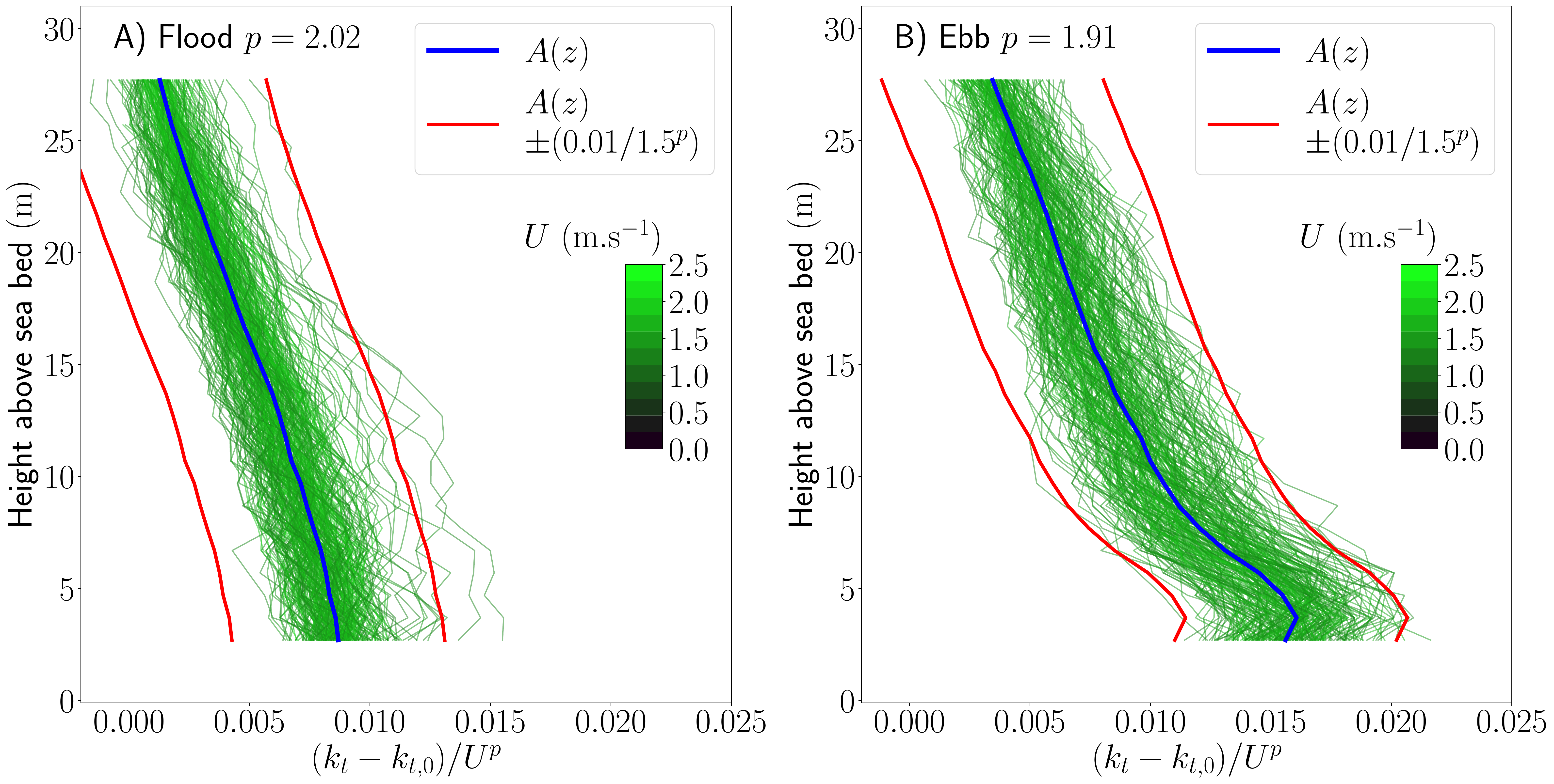}
 \caption{Scaled TKE profiles $\left(k_t(z) - k_{t,0}(z)\right)/{U}^p$ for flood (plot A) and ebb (plot B) cases coloured with respect to the barotropic velocity, only samples with $U>1.5\,\si{\m\per\s}$ are kept, low velocity profiles are showing too much divergence; the blue profile corresponds to the shape function and scaled TKE $A(z)$, the red profiles correspond to an upper-bound estimation of the error induced by the variability coming from remnant unsteadiness still present in the TKE data}
 \label{tke_scaled}
\end{figure}

As stated previously in Section \ref{section_method}, the first step is to verify that for each tidal phase a scaling by the velocity $U$ to some power is possible. This requires that the power fits for $\alpha$ and $\beta$ in Eq. (\ref{reg_1}) give roughly a same estimation for $p_\alpha$ and $p_\beta$. This is indeed verified in Fig. \ref{alpha_beta} with a relative difference of less than $5\%$ for the ebb cases, they are slightly higher for the flood cases with $9\%$. All values are around the theoretical expectation $p=2$. The agreement for the ebb cases is probably incidental especially given the dispersion observed for the $\alpha$ points. The mean values as per Eq. (\ref{reg_2}) are kept for each tidal phase.

The bin per bin regressions are then carried out to evaluate $A(z)$ and $k_{t,0}(z)$. The latter are shown with Fig. \ref{tke} featuring the raw profiles. In each case $k_{t,0}(z)$ is indeed small compared to the profile, especially near the bottom. However it can be of similar order of magnitude further up in the water column with maximum values around $0.01\,\si{\square\m\per\square\s}$.
The value is reassuring as it is of the same order and in general less than the errors carried by the TKE and attributed to the unsteadiness of the flow. The slight increase further from the bottom could also be a signature of surface and wave effects already impacting the turbulence or advected turbulent structures generated prior to the ADCP. None have a reason to scale with the barotropic velocity.

The general shape of the profiles is well captured by the regression with the $A(z)$ coefficient (Fig. \ref{tke_scaled}). We decide at this stage to remove all the profiles with low $U$ as they yield diverging scaled TKE profiles due to the scaling. A safe threshold is picked as we only keep profiles with $U>1.5\,\si{\m\per\s}$ in order to stay consistent with the rest of the analysis.
Visually strong differences are observed between ebb and flood, which can be explained by different bathymetric features encountered upstream of the ADCP in each case, as observed in \cite{mercier2021impact}. The peak value and strong shear for ebb cases near $5\,\si{\m}$ above seabed is also observed in \cite{mercier2021turbulence}, although they mention that it could be an measurement artefact.
The collapse of the TKE profiles is not perfect, a lot of dispersion still appears on the scaled individual profiles which is a direct consequence of the variability carried by the data already mentioned. The error estimate is large compared to the mean scaled profile $A(z)$, the uncertainty can reach up to $66\%$ of the estimated value. Indeed, we compute a rough upper-evaluation of the model uncertainty using the value $0.01\,\si{\square\m\per\square\s}$ scaled by the lowest velocity $U=1.5\,\si{\m\per\s}$ to the appropriate power $p$. However, we still expect large confidence intervals would they be computed more precisely.

\begin{table}
\begin{center}
\begin{tabular}{l c  c  c | c  c  c}
\toprule
 & \multicolumn{2}{l}{$0\,\si{\m\per\s}<U<2.8\,\si{\m\per\s}$} & & & \multicolumn{2}{l}{$1.5\,\si{\m\per\s}<U<2.8\,\si{\m\per\s}$} \\
 & RMSE & $R$ & & & RMSE & $R$              \\
 & ($\si{\m^2\per\s^2}$) & & & & ($\si{\m^2\per\s^2}$) &   \\
\midrule
Flood        & $0.0034$ & $0.63$ & & & $0.0039$ & $0.90$ \\
Ebb          & $0.0041$ & $0.69$ & & & $0.0050$ & $0.94$ \\
\bottomrule
\end{tabular}
\end{center}
\caption{RMSE and correlation coefficient $R$ for the TKE model defined with Eq. \ref{kt_model}; averaged over all profiles on the left part of the table while only over cases with $U>1.5\,\si{\m\per\s}$ on the right part; the biases are not shown as close to $0$ in both cases.}\label{stats_tke}
\end{table}

With the parameters $A(z)$, $k_{t,0}(z)$ and $p$ evaluated, it is now possible to have an estimate of the base TKE generated by the tidal currents through their interaction with the bottom as long as the tidal barotropic velocity is known, using Eq. (\ref{kt_model}). The accuracy of the model is given in Table \ref{stats_tke}, featuring both all $U$ cases and cases with $U>1.5\,\si{\m\per\s}$ only.
The RMSE are slightly smaller when including all cases ($0.0034-0.0041\,\si{\m^2\per\s^2}$ against $0.0039-0.0050\,\si{\m^2\per\s^2}$) where we know that the shape of the TKE profiles are badly captured by the model. This is not surprising as the TKE is then generally smaller, meaning the amplitude of the dispersion is also smaller in absolute value. The correlation is however bad ($0.63-0.69$) due to the shape not well captured by the model. For those low velocities other mechanisms then friction of the tidal currents at the bottom can compete and deviate from the expected shape. This is the same distinction previously observed in Fig. \ref{vel_profiles} where the velocity profiles do not agree with the theoretical logarithmic profiles expected.
On the other hand for well-developed states with strong velocities the correlation increases drastically ($0.90-0.94$). This is the signature of the scaled TKE collapsing well on the shape parameter $A(z)$.
Ebb cases carry a stronger error, but a better correlation. This can be partially explained by the difference in absolute TKE values between the two tidal cases. Ebb cases are in general more energetic than flood cases (Fig. \ref{tke}), therefore the absolute errors tend to be stronger while the relative correlations better. 

\subsection{Discussion}

\begin{table}
\setlength{\tabcolsep}{5pt}
\begin{center}
\begin{tabular}{l l  c c c l c c c}
\toprule
Scaling &        & $U^p$  &  & $u_\tau^2$   &      & $U^p$   &   & $u_\tau^2$ \\
\midrule
Normalised STD   & Flood  & $28\%$ & $<$ & $41\%$    & Ebb  & $20\%$ & $<$ & $33\%$ \\
\bottomrule
\end{tabular}
\end{center}
\caption{Normalised standard deviation of the scaled TKE with either $U^p$ or $u_\tau^2$, for $U > 1.5\,\si{\m\per\s}$.}\label{std_tke}
\end{table}

A direct example of the model benefits is illustrated with Table \ref{std_tke}, where the normalised standard deviation of the TKE scaled with either $U^p$ or $u_\tau^2$ are compared. We thus evaluate the scaling obtained with the model introduced in this paper against the theoretical expectations from the wall theory. We find that for both flood and ebb cases our model is reducing the normalised deviation, $28\%$ against $41\%$ for flood cases and $20\%$ against $33\%$ for ebb cases. This is indicating that the model and method presented here is performing better than the theory.

As already mentioned the velocity model, we propose with Eq. (\ref{v_model}) is unable to predict correctly the velocity profiles. The errors are linked to a bad estimation of the roughness length which should not be taken constant, and lead to strong biases between the model and measurements. However since the error is mostly a strong bias it is still safe to use the model when the velocity shear is needed, which will be done in the next section.
A possible issue with the method is the lack of measurements closest to the seabed, where the roughness is actually at play. The profiles are fitted over the whole water column and as a result the near-bed part, crucial in estimating correctly the roughness height, is not necessarily well captured.
This limitation is inherent to the use of ADCPs due to the blanking region of roughly $2\,\si{\m}$ between the first bin and the transducer head \cite{togneri2017comparison,lewis2017characteristics}. However it can be overcome with the use of an Acoustic Doppler Velocimeter (ADV). For instance \cite{feddersen2007direct} develop and test a structure with three ADVs and the first one at $0.56\,\si{\m}$ above the sea bed.
ADVs also measure at a higher frequency, from $8\,\si{\Hz}$ to $12.5\,\si{\Hz}$, which is $4$ to $6$ times the frequency of a classic ADCP. As such the velocity spectra and consequently the TKE and dissipation are better estimated. In our case ADV data is also available \cite{furgerot2020measurements}, but the data presents quality issues and since the record time and location of the ADV don't match exactly those of the ADCP used here we decide to discard this set of data.

Although Eq. (\ref{kt_model}) only requires the barotropic velocity to provide an estimate for the tidal-generated TKE, it is still necessary to measure and collect the raw beam velocities from the ADCP in order to estimate the model parameters $A(z)$, $k_0$ and $p$. Furthermore it has to be done during a season of calm sea states in order to have data unaffected by waves. This is not a limitation of the method per-say, but it can make its applicability less affordable and is worth mentioning. Moreover the authors do not recommend the use of generic values, especially for the $A(z)$ parameter. As shown in Fig. \ref{tke_scaled}, the shape can vary greatly from one flow condition to another, and certainly from one location to another as observed in \cite{bourgoin2020turbulence}. The surrounding bathymetry is most certainly the cause for this variability as stated in \cite{mercier2021impact}, but this aspect is not studied here.

We have also mentioned but neglected so far the impact of the tidal phase, although it has been shown to be relevant \cite{togneri2016micrositing}. One given $U$ value can correspond to two different configurations, either when the tidal currents are increasing at the beginning of the tidal flow, or when the tidal currents are decreasing. This is inducing slight trends in the data, they are deemed small enough that they can be neglected.

A more direct approach to evaluate the TKE profiles is possible by directly conducting brute optimisations bin per bin, estimating all three parameters $A(z)$, $k_{t,0}(z)$ and $p(z)$ at the same time. The initial scaling step where an estimate for $p$ is first found for the entire water column is then skipped. Although it should give a better fit in theory since it allows for more degrees of freedom, it is not preferred since we lose the possibility to scale the TKE uniformly as the power exponent $p(z)$ would then vary with depth.

%----------------------------------------------------------------
\section{Applications \label{section_applications}}
%----------------------------------------------------------------

The models given by Eqs. (\ref{log}) and (\ref{kt_model}) are first tested to the evaluation of the TKE budget terms. We compare the prediction from the models to measurements of the budget terms. The results are commented notably with references to the wall theory. 
A second application is carried out to estimate and characterise the wave-induced velocity and TKE.

\subsection{Budget terms and departure from the wall theory}

\begin{figure}
 \centering \includegraphics[height=7cm]{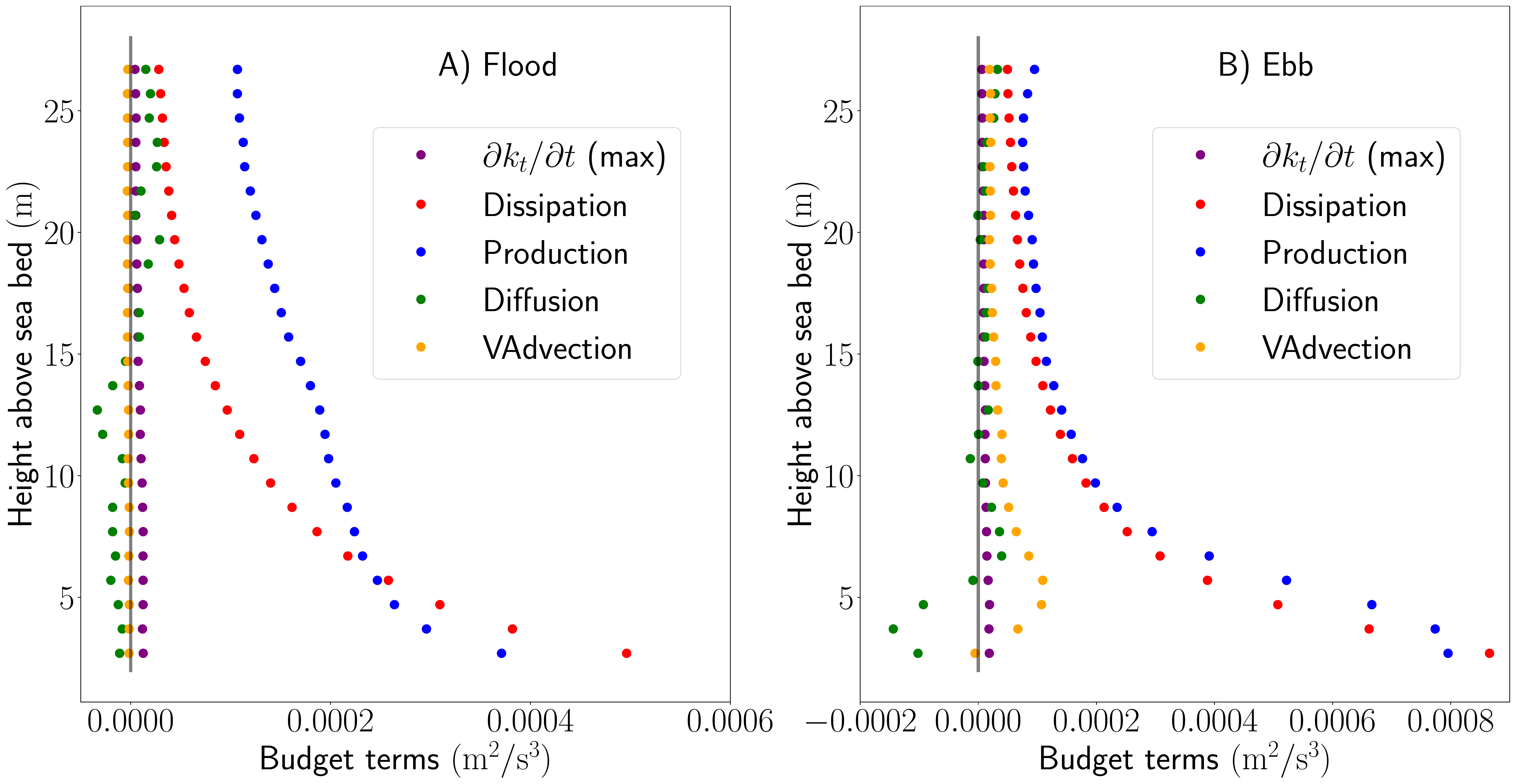}
 \label{budget}
 \caption{Averaged profiles of the budget terms in Eq. (\ref{tke_balance}) for flood (plot A) and ebb (plot B) cases; the maximum absolute value for the time variation of the TKE is shown instead of the mean value.}
\end{figure}

A first direct assessment of the budget terms appearing in Eq. (\ref{tke_balance}) is conducted on each $20\,\si{\min}$ sample featuring a calm sea state. Direct measurements of the mean velocity, TKE, as well as Eqs. (\ref{lm}, \ref{epsilon_kt}, \ref{nu_kt}) are used to this effect. The mean values are shown in Fig. \ref{budget}, which highlights well that the dominant terms are the turbulent dissipation $\epsilon$ and production $\mathcal{P}$.
We then decide to focus solely on those two terms and neglect the temporal variation, vertical advection and diffusion of TKE.

Using Eqs. (\ref{lm}, \ref{log}, \ref{kt_model}) we find the following expressions for $\mathcal{P}$ and $\epsilon$, scaled with the barotropic velocity $U$. For simplicity we decide at this stage to neglect the error term $b_\tau$ when modelling the friction velocity. We also make the assumption that the direction of currents is roughly constant over depth. It is a strong modification not necessarily verified \cite{fugerot2018velocity}, even during the peak of the tide, but it allows to introduce the derivative of the current magnitude instead of retaining the two horizontal directions:

\begin{align}
& \mathcal{P} = c k_t^{1/2} l_m \left(\dpar{u}{z}\right)^2  =
    c a_\tau^2 {\kappa} \frac{(A + k_{t,0}/U^p)^{1/2}}{\kappa z} U^{(p/2) +2}, \label{production}\\
& \epsilon = c_\epsilon k_t^{3/2}/l_m = c_\epsilon \frac{(A + k_{t,0}/U^p)^{3/2}}{\kappa z} U^{3p/2} . \label{dissipation}
\end{align}

As a first comment the scaling between the production $\mathcal{P}$ and the dissipation $\epsilon$ is a-priori different, $(p/2 +2)$ against $(3p/2)$ respectively. A similar scaling is found if and only if $p=2$, which is the theoretical expected value and almost found by the method and regressions shown with Fig. \ref{alpha_beta}.
For the ebb cases a good balance is observed between production and dissipation with those expressions, however it is not the case for the flood cases.
A possible source of error from Eq. (\ref{tke_balance}) is the omission of the horizontal advection of TKE, which is neglected as the theory relies on a horizontally homogeneous flow. It is however impossible to estimate this budget term as we only have data available at one location.
The validity and accuracy of each budget term will studied further in the next sections.

\subsubsection{TKE production}

\begin{figure}
 \centering \includegraphics[height=7cm]{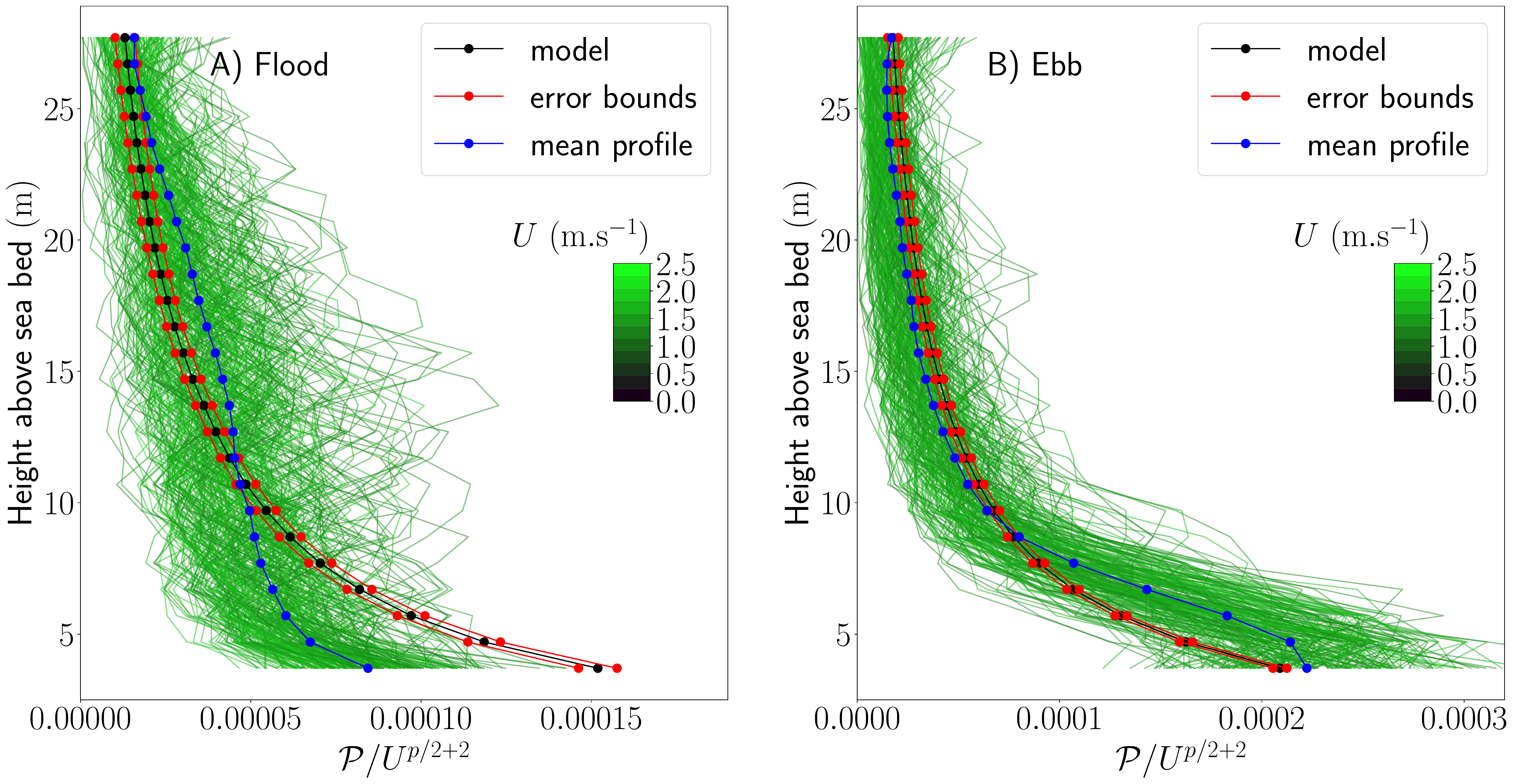}
 \caption{Production $\mathcal{P}$ scaled with $U^{p/2+2}$ for flood (plots A) and ebb cases (plots B); the green coloured curves correspond to the profiles estimated from TKE and velocity measurements as well as mixing length model, the black curves are the estimates using the velocity, TKE and mixing length models given by Eqs. (\ref{lm}, \ref{log}, \ref{kt_model}), the red curves are the model estimate error bounds; only profiles  with $U > 1.5\,\si{\m\per\s}$ are kept.}
 \label{budget_production}
\end{figure}

The scaled expression on the right part of Eqs. (\ref{production}) is now compared against its direct estimation on the left part of the equation, in Fig. \ref{budget_production}. The green curves show the direct estimation profiles, involving the TKE, the mean velocity vertical derivative and the mixing length $l_m$ modelled with Eq. (\ref{lm}). The black curves correspond to the model scaled expression where Eqs. (\ref{lm}, \ref{log}, \ref{kt_model}) are used, bounded by error estimations with the red curves.
We observe strong discrepancies for the turbulent production, especially near the bottom.

The main issue lies in a correct estimation of the velocity shear. The vertical resolution of the data is rather crude ($1\,\si{\m}$) which introduces some uncertainty in the direct estimation. But most importantly, as already observed with Fig. \ref{vel_profiles}, the logarithmic profile fails to capture the velocity shear correctly through the whole water column. It leads to substantial errors in the model estimation of the production term, which seems to be more important near the bottom.

\subsubsection{TKE dissipation}

\begin{figure}
 \centering \includegraphics[height=7cm]{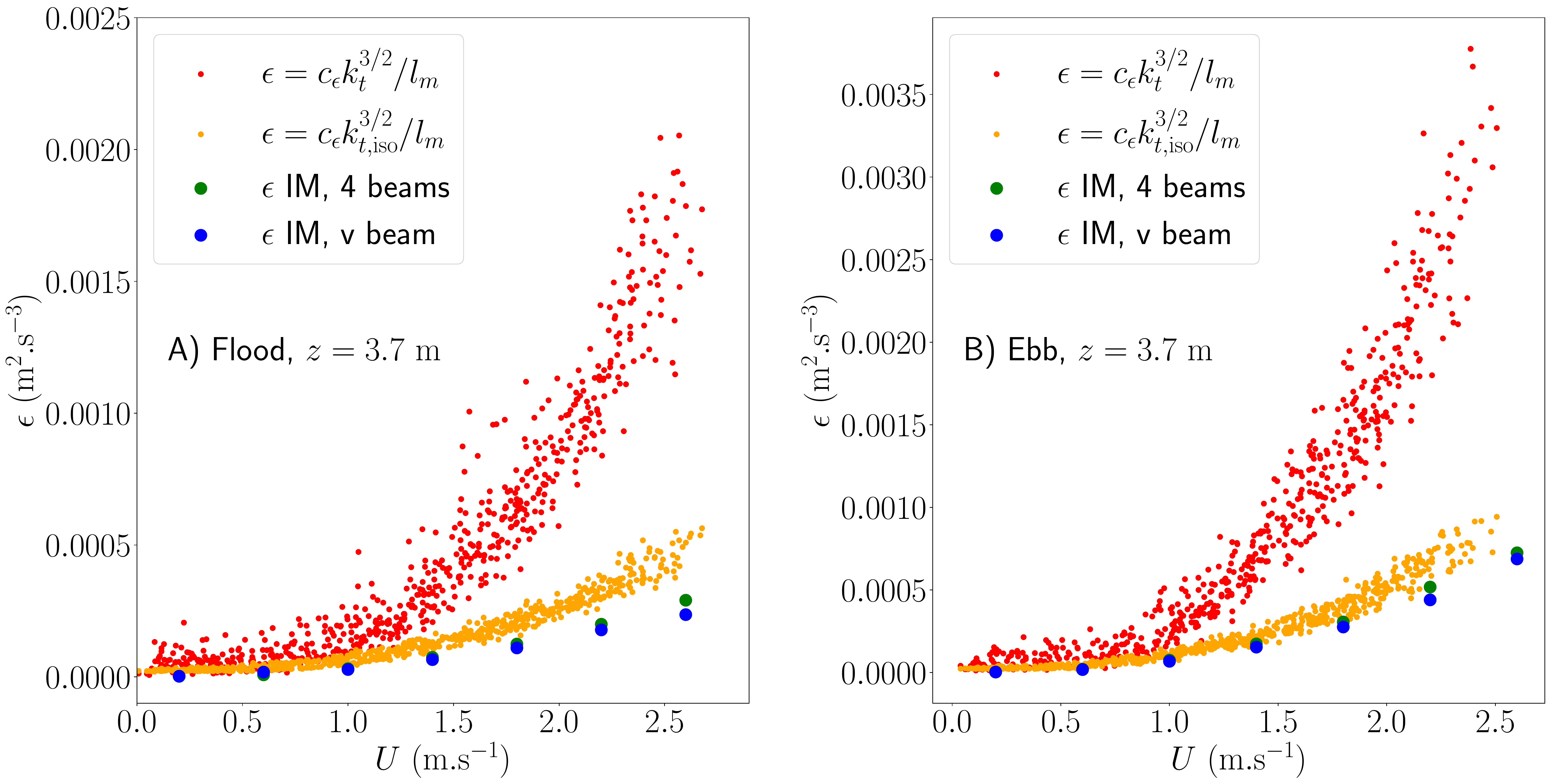}
 \caption{Turbulent dissipation evaluated close to the seabed at $z=3.7\,\si{\m}$, for flood (plot A) and ebb (plot B) cases, the green and blue dots are direct measurements using either the integral method on the $4$ inclined beams or the vertical beam respectively, the red dots are estimations using the measured turbulence and the theoretical relation Eq. \ref{epsilon_kt} while the orange dots use the corrected isotropic estimation of turbulence $k_{t,\text{iso}}$.}
 \label{dissipation_compare}
\end{figure}

We have direct and independent measurements available for the turbulent dissipation using the integral method described in Section \ref{integral_method}. We compare those to the estimate from the wall theory given by Eq. (\ref{epsilon_kt}) in Fig. \ref{dissipation_compare}. We show on purpose a level close to the bottom ($z=3.7\,\si{\m}$), where the wall theory is more likely to hold. Despite it a considerable discrepancy is observed between the direct measurements in blue and green dots and the estimate from the TKE in red. The integral method requires to average the spectra over velocity bins, which is why we do not have a dense scatter plot but instead average estimates around each velocity bin.
The two integral methods are in good agreement, with values reaching at most  $0.0007\,\si{\square\m\per\cubic\s}$, but the estimates using Eq. (\ref{dissipation}) are almost $3$ times larger with maximum values around $0.004\,\si{\square\m\per\cubic\s}$. Both ranges of values are reasonable though, and can be found in the literature \cite{mcmillan2016rates,thiebaut2020comprehensive}.
A direct consequence of this observation is that Eq. (\ref{dissipation}) is not valid and should not be used to obtain a correct estimate of the dissipation.

This is a major departure from the wall theory. In a real oceanic application the turbulence is in general anisotropic at large enough scales, with highly energetic horizontal structures. However the theoretical relationship between the dissipation and the TKE given by Eq. (\ref{epsilon_kt}) is derived and verified assuming an purely isotropic turbulence. It is therefore expected to overestimate the dissipation if anisotropic structures are included in the estimation of $k_t$.
In order to correct it, we expect that the vertical normal stresses $\overline{w'^2}$ are mostly including small scales where turbulence is isotropic, which leads to an evaluation of the isotropic turbulence with $k_{t,\text{iso}} = \frac{3}{2} \overline{w'^2}$.
The dissipation estimated from $k_{t,\text{iso}}$ is shown in the same Fig. \ref{dissipation_compare} with orange dots. The agreement is significantly improved although the dissipation is still overestimated when using the theoretical formula. It is likely that even $\overline{w'^2}$ carries larger anisotropic structures leading to this overestimation. 

\subsubsection{Updated budget balance}

\begin{figure}
 \centering \includegraphics[height=7cm]{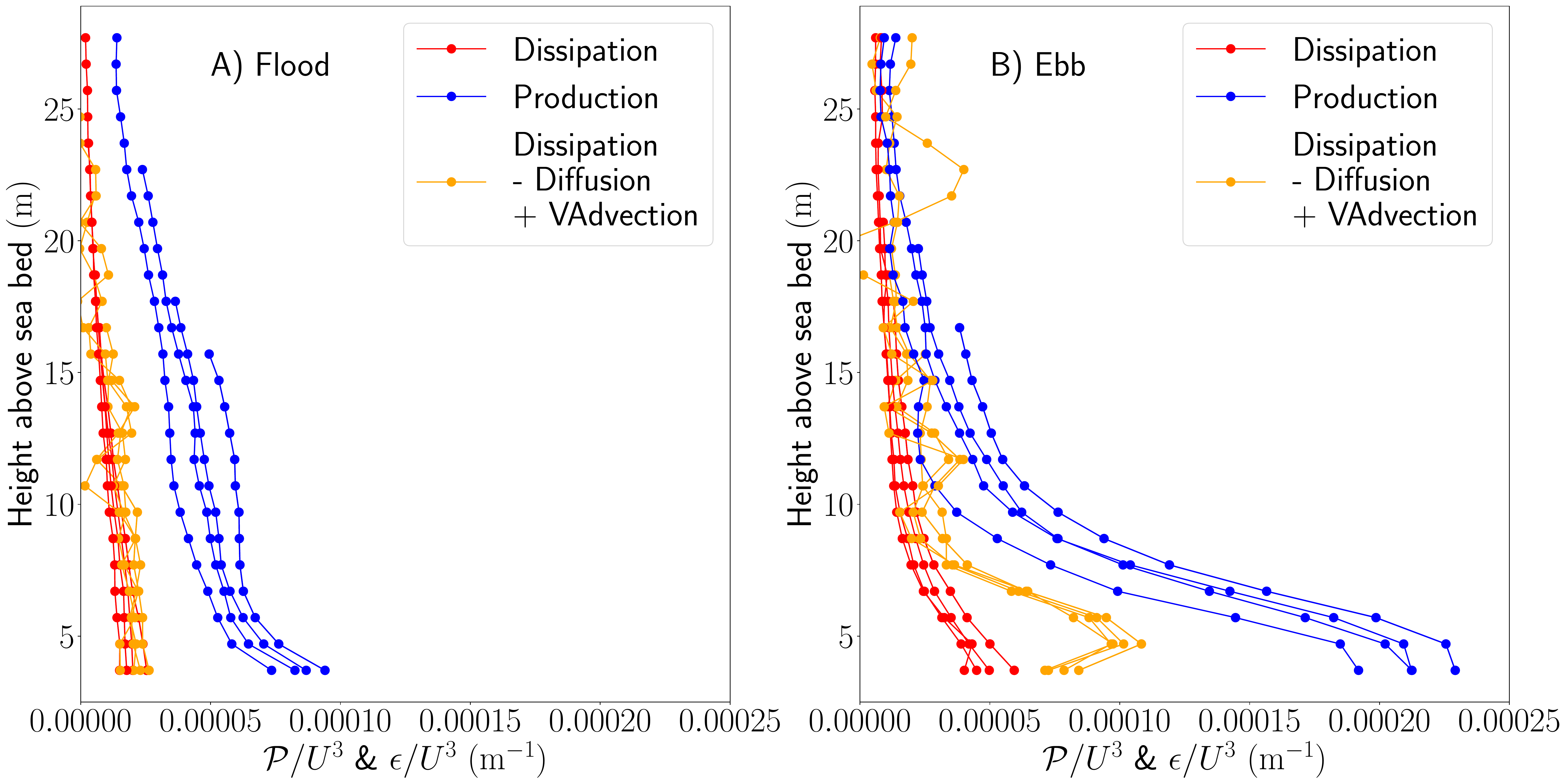}
 \caption{Scaled profiles of the turbulent production (blue), dissipation (red) and three main budget terms matching the production (orange) for flood (plot A) and ebb (plot B) cases; direct measurements of the turbulent kinetic energies, dissipation and mean velocities are used and averaged around $U$ velocity bins of width $0.4\,\si{\m\per\s}$, only the $4$ velocity bins above $1.4\,\si{\m\per\s}$ are shown for each budget term.}
 \label{budget_updated}
\end{figure}

The prediction models presented in this paper are enable to reproduce correctly the production and dissipation budget terms using Eqs. (\ref{production}, \ref{dissipation}). The shape of production budget term is not well reproduced due to the velocity shear badly captured by the log of the law, especially near the bottom. For the dissipation the relationship used with Eq. (\ref{epsilon_kt}) is not applicable due the anisotropic characteristic of the TKE.
The purpose of this paragraph is to provide a faithful picture of the budget balance. Since the scaling exponent $p$ is close to $2$ in both flood and ebb cases for the present application we will assume this value and use it to scale the production and the dissipation by $U^3$. The budgets are computed using the direct measurements for both the TKE, dissipation and mean velocity, and shown in Fig. \ref{budget_updated}.

The scaling and collapsing of the budget is not perfect, but sufficient to notice strong differences between production and dissipation, with different behavior according the tide.
Overall and as expected from the previous observations the dissipation is way smaller than the production, and more intensity in both processes is observed during ebb cases.
The gap is almost constant during flood phases with production and dissipation profiles almost parallel to each other, the production is roughly $3$ times stronger than the dissipation. On the other hand for ebb cases a good balance is observed above $15\,\si{m}$ depth, with only a clear discrepancy appearing closer to the seabed. The production can then reach $4$ times the dissipation.

The discrepancy between production and dissipation is unexpected as the wall theory suggests a balance between those two processes. The neglected mechanisms mentioned earlier and shown in Fig. \ref{budget} are not important enough to close completely the gap, although they reduce it considerably for ebb cases. For both flood and ebb the unbalance in the budget terms is of the order $8 U^3\,\si{\m^2\per\s^3}$.
The theory assumes a horizontally homogeneous flow which has no reason to be verified here in a real application. We are lacking data to assess how the flow characteristics might vary spatially, horizontal advection could still play an important role.
Another argument in favour of this explanation is the similarity in shear observed for the velocity between flood and ebb, highlighted with plots (A, B) in Fig. \ref{log_coef}. As such the difference in turbulent production has to come for the turbulent viscosity, and by extension the TKE itself. The only remaining source of TKE is then through horizontal advection.
For the ebb cases this supposed horizontal advection would appear near the seabed, and trigger a strong vertical advection and diffusion. This behavior would agree with eye observation of the area where large eddies are observed to surface even by calm sea states \cite{mercier2020numerical}.
The data presented in Fig. \ref{budget_updated} would suggest a repetitive and localised pattern, but there is sadly not specific studies on those emerging eddies to verify or deny this explanation.
What the authors struggle to explain is the constant unbalance during flood cases, especially above $z=15\,\mathrm{m}$. Compared to the ebb cases it seems to be linked a smaller dissipation, but this difference in behavior is not explained.

\subsubsection{Departure from the wall theory}

The wall theory has been used as a reference extensively in this paper, in order to comment on the different results obtained either directly with the estimation of the parameters in Eq. (\ref{kt_model}), or with the study of the budget terms. However the conclusions are that in general a large departure from the wall theory is observed.

The shape of the measured TKE is hardly showing the expected shape obtained from experiments or DNS, a layer of constant TKE \cite{krogstad2005experimental,chatzikyriakou2015turbulent} followed by a linear decay should be observed but is not obviously reproduced by the shape function $A(z)$ shown in Fig. \ref{tke_scaled}. We even identify for ebb cases an initial increase in TKE with a maximum around $5\,\si{\m}$ above seabed.
A similar but smaller departure is observed in \cite{mercier2021turbulence}, they conduct a large-eddy simulation of a high Reynolds flow over a rough seabed where the roughness at the bottom is physically modelled by an array of $3\,\si{\m}$ large cubes. Their flow conditions are more similar to the theory and yet they still observe transverse Reynolds stresses departing from their theoretical profiles, notably lacking a clear constant shear layer.
In return, this departure from the theory questions the validity of the law of wall which is used here, but also in many oceanic applications. The fact remains that it gives a good agreement for the velocity, but as observed when estimating turbulent production the velocity shear is not well captured by the law, especially near the seabed.
Anyhow, this first point is a strong argument that another process than friction is relevant and generate turbulence near the seabed in our application characterised by strong tidal currents.

Through direct comparisons it is observed that the theoretical formula given by Eq. (\ref{epsilon_kt}) is invalid. The hypothesis explored is that the strong anisotropy of turbulence is causing an overestimation of the dissipation, that should only be evaluated in the isotropic range.
Anisotropy is common and expected for oceanic applications, large scale horizontal eddies have been observed profusely in the literature (e.g. \cite{stacey1999measurements,nezu1993turbulence,thomson2012measurements,milne2013characteristics}), but with no impact either on the direct TKE or turbulent dissipation measurements nor on the validity of the law of the wall. 
In particular they compare in \cite{stacey1999measurements} the friction velocity obtained from direct fit to the data against an estimation from the Reynolds stresses, they find a good agreement.
Using Eqs. (\ref{ut_kt}, \ref{kt_model}) and the direct fit shown with Fig. \ref{log_coef} we indeed verify this good agreement in our application with $c \sqrt{A} \approx 0.06 \pm 0.01$, neglecting the impact of $p$ not being exactly $2$.
As a result the authors believe that anisotropy is only putting into question Eq. (\ref{epsilon_kt}), usually introduced by dimensional analysis in the isotropic range, and not the other terms in the budget balance given by Eq. (\ref{tke_balance}).

\subsection{Wave-induced velocity and TKE}

\begin{figure}
 \centering \includegraphics[height=7cm]{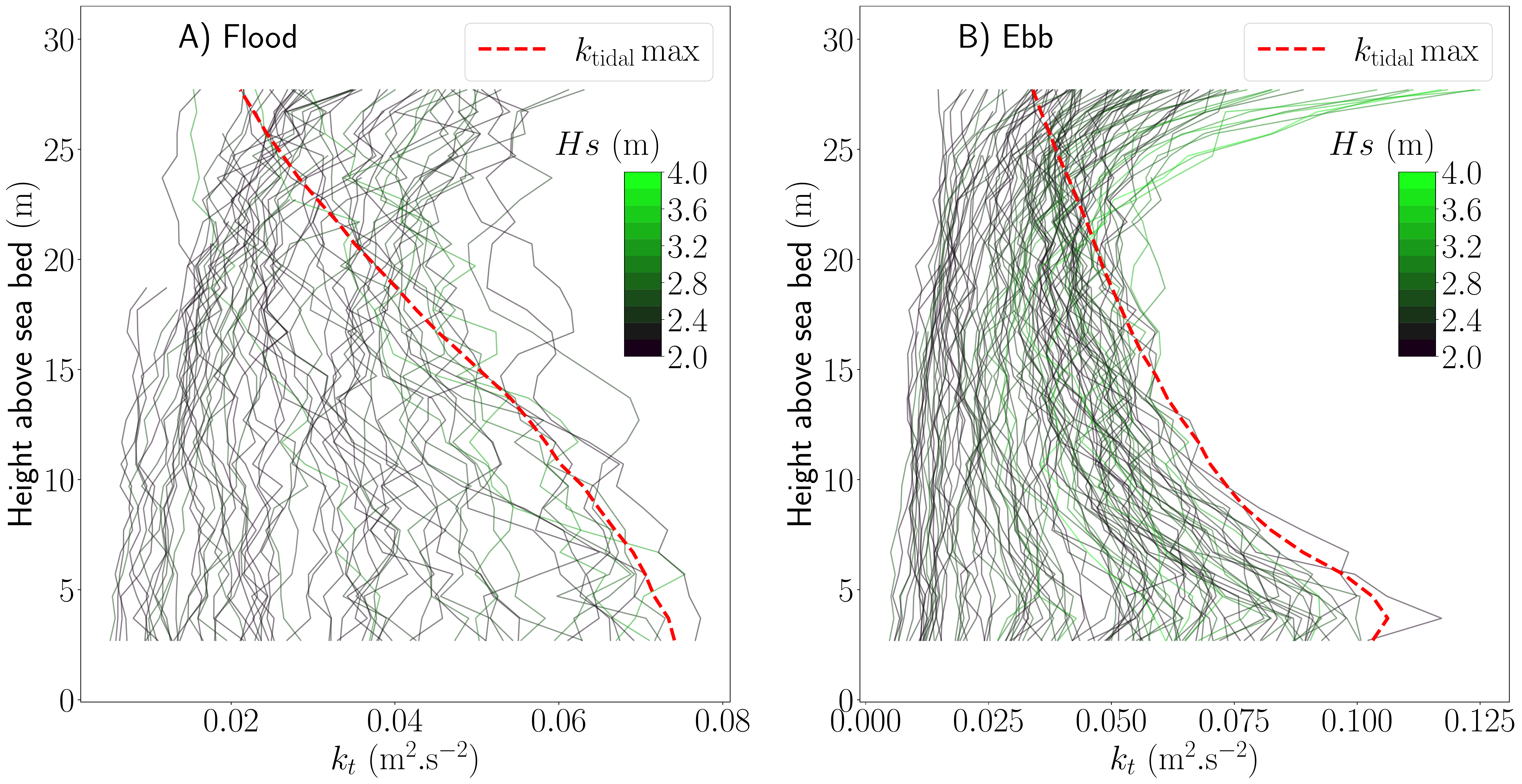}
 \caption{Turbulent kinetic profiles for flood (plot A) and ebb (plot B) cases showing only samples with $H_s > 2 \,\si{m}$, profiles are coloured by the significant wave height and the red dot curve shows the maximum profile generated by the tidal currents.}
 \label{tke_wave}
\end{figure}

At last we now want to use the fitted models given by Eqs. (\ref{v_model}, \ref{kt_model}) in order to retrieve the wave contribution to the mean velocity and TKE profiles. 
The impact of waves and the necessity of removing the tidal-generated part is easily observed on the TKE shown in Fig. \ref{tke_wave}. It mimics the previous Fig. \ref{tke} except that only sea states with $H_s>2\,\si{\m}$ are kept, and the profiles are coloured by the significant wave height.
To ease the comparison the maximum tidal-generated TKE profile is shown with the red dot curve. The impact of waves on the turbulence is immediately observed in the upper levels, down to $20\,\si{\m}$ above the seabed. Below this value the tidal-generated component always dominates.

\begin{figure}
 \centering \includegraphics[height=7cm]{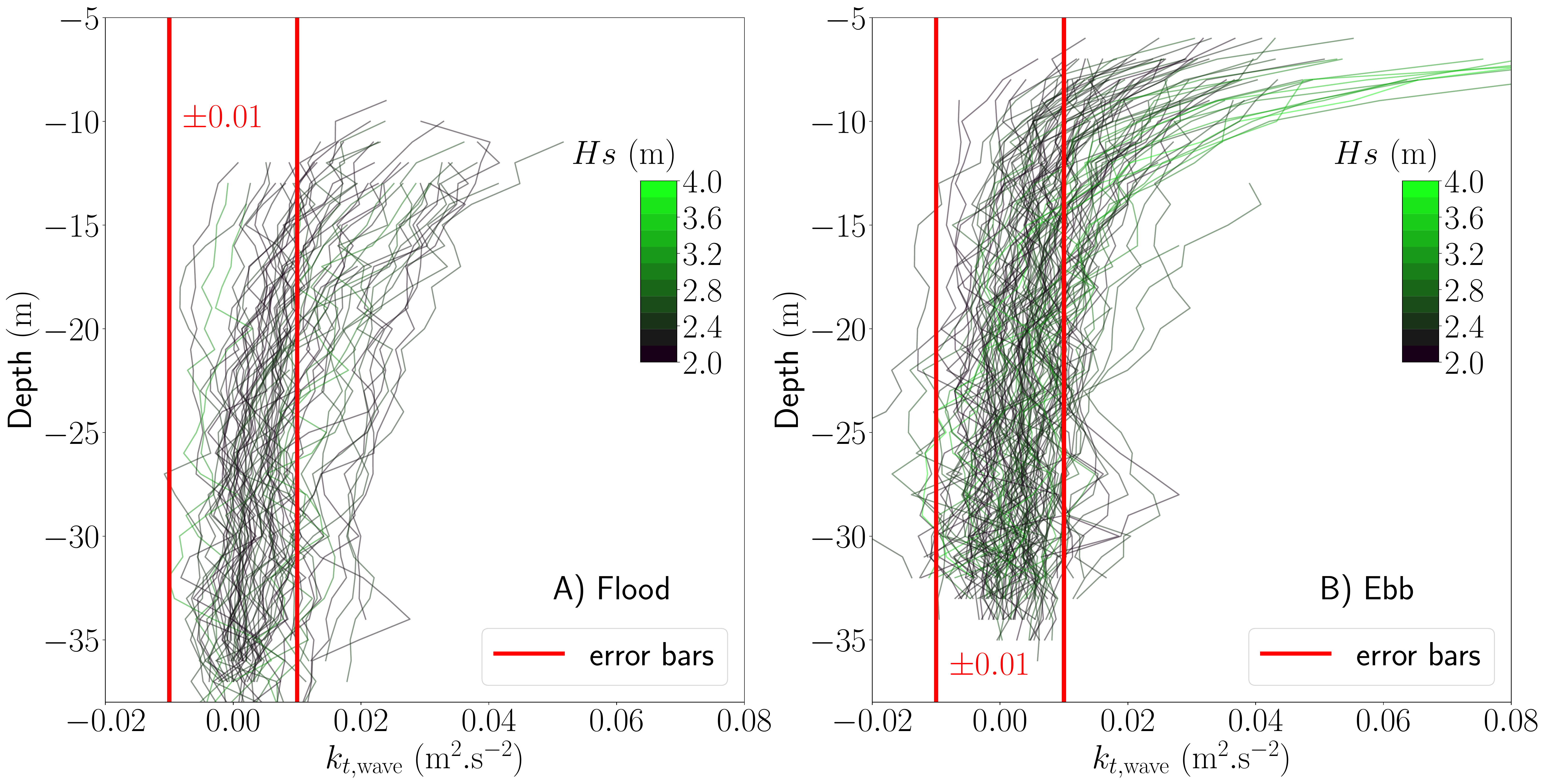}
 \caption{Wave-induced turbulent kinetic profiles for flood (plot A) and ebb (plot B) cases showing only samples with $H_s > 2 \,\si{m}$, profiles are coloured by the significant wave height and the vertical axis is now the depth; the red bars indicate the uncertainty carried by the data (see Fig. \ref{vel_tke_error}).}
 %added to the prediction model error $k_{t,0}(z)$.}
 \label{tke_wave_inv}
\end{figure}

Looking now specifically at the wave-induced profiles with Fig. \ref{tke_wave_inv}, we observe a clear scaling with the significant wave height. For consistency between the profiles the vertical axis has been reversed to show the depth instead of the height above seabed, the distance from the sea surface is indeed preferred since we are now studying the impact of waves. The difference in position between the flood and ebb cases is directly linked to the tidal sea surface variation, and it is actually critical to observe agreement between the flood and ebb profiles.
We will only carry a qualitative analysis. From the ebb cases we notice that the strongest profiles all correspond to events with large significant wave heights, as expected. It is not easy to interpret the core of the profiles since the impact of waves does not seem to reach deep enough in the water column with amplitudes close to the estimate for the measurement errors. For simplicity we decide to use this estimate for the error instead of a more accurate model uncertainty analysis not carried out here.
This is certainly the stronger limitation of the method presented in this paper relying on bottom-mounted ADCP data, the first $6\,\si{\m}$ below the surface can never be exploited which is unfortunate in order to observe wave effects.

\begin{figure}
 \centering \includegraphics[height=7cm]{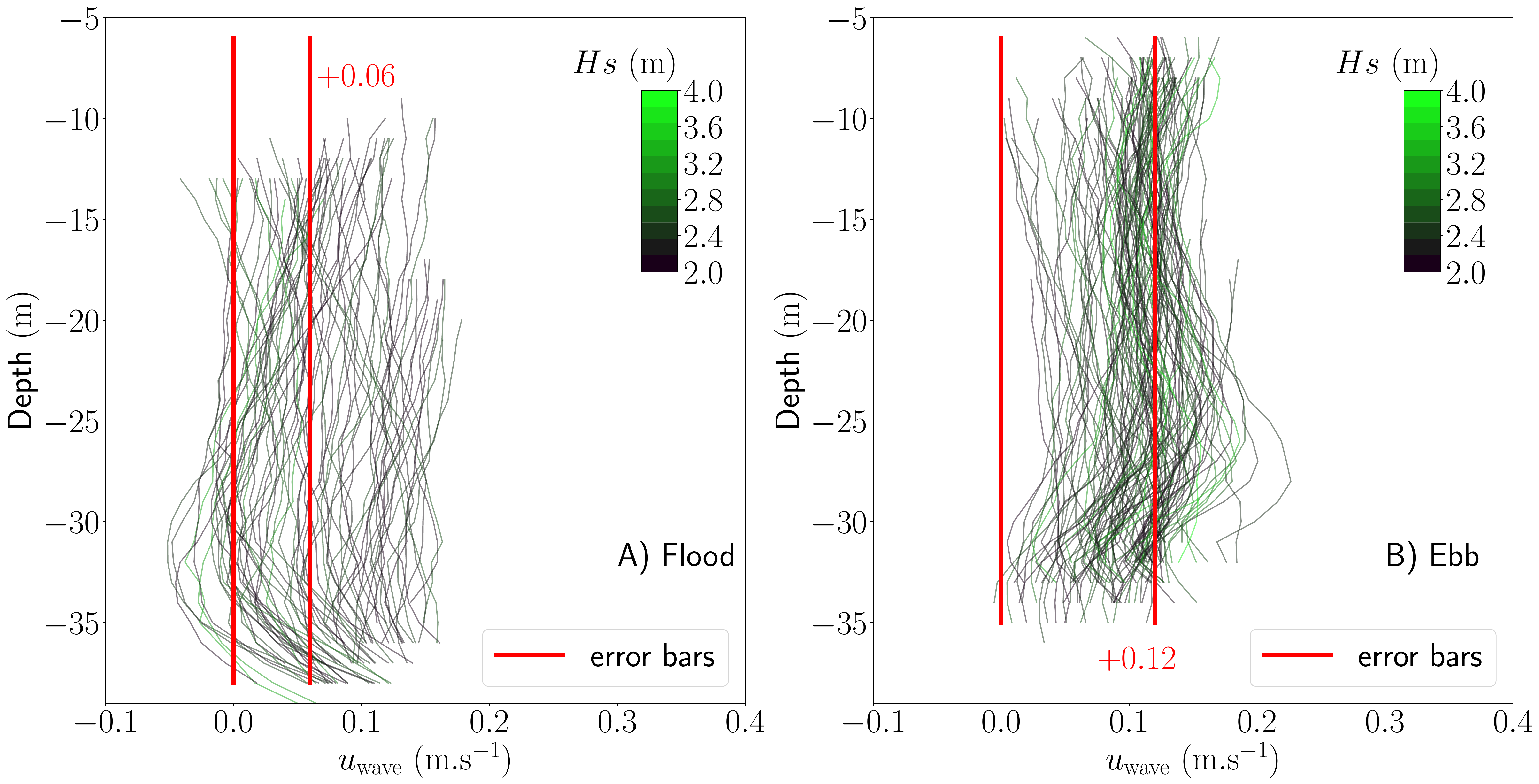}
 \caption{Wave-induced velocity magnitude for flood (plot A) and ebb (plot B) cases showing only samples with $H_s > 2 \,\si{m}$, profiles are coloured by the significant wave height and the vertical axis is now the depth; the red bars indicate roughly the model errors using the bias (see Table \ref{stats_vel}).}
 \label{vel_wave_inv}
\end{figure}

The impact of waves on the velocity profiles is shown in Fig. \ref{vel_wave_inv}. 
%We are unable to conclude anything due to the strong bias error carried by the model. Those errors are measured at $0.06\,\si{\m\per\s}$ for flood cases and $0.12\,\si{\m\per\s}$ for ebb cases (see Table \ref{stats_vel}). Either way they are of similar magnitude than the variance observed in the wave-induced velocity profiles.
%Moreover, even if we decide to ignore the error uncertainty, there is no obvious impact of the significant wave height on those profiles, which is a dependence that we would expect regardless of the dominant wave-current process.
It is hazardous to conclude anything from those profiles for two reasons. First of all the strong bias error carried by the model, $0.06\,\si{\m\per\s}$ for flood cases and $0.12\,\si{\m\per\s}$ for ebb cases (see Table \ref{stats_vel}), are of similar magnitude than the variance observed in the wave-induced velocity profiles. Additionally there is no obvious impact of the significant wave height on those profiles, which is a dependence that we would expect regardless of the dominant wave-current process. A limiting factor is the variability in the wave direction relatively to the current direction. As observed in canal experiments and verified with numerical models \cite{groeneweg1998changes,olabarrieta2010effects} a decrease in mean velocity shear is expected for following wave and tidal current, along with a reduction of amplitude in the upper-half and a small amplification in the lower-half. It is the other way around in case of opposing wave and tidal current. Finally for perpendicular wave and tidal-current an increase of shear is observed with an intensification of the mean velocity.
This would require additional analysis not carried out in the present paper.

\section{Conclusion}

The formula introduced with Eq. (\ref{kt_model}) and method proposed to evaluate the different parameters is able to provide a simple prediction model only depending on the barotropic velocity. It evaluates accurately the TKE generated by the tidal currents. However we could not obtain a reliable model for the velocity, Eq. (\ref{v_model}) is introduced but the roughness height is badly estimated with no clear dependence towards the barotropic velocity. It results in bias errors of the order of $0.1\,\si{\m\per\s}$ at worst, which might be too much for refined power resource assessment, or for estimating the wave-induced contribution to the velocity profiles.
A strong limitation not explored here is the lack of data points near the seabed. It is an operational limitation when using only one bottom-moored ADCP, but it could be overcome with the use of an ADV for instance.

We use the two models for an initial study of the budget terms during calm sea states. However the analytical formulas derived from our models for both the production and dissipation, given by Eqs. (\ref{production}, \ref{dissipation}), both fail at evaluating correctly those budget terms. For the production term the main issue lies in a bad estimation of the velocity shear. A correct estimate overall is obtained but the analytical prediction fails at capturing accurately the near-bed production. As for the dissipation, direct measurements are available from the ADCP data and we find that the theoretical formula Eq. (\ref{epsilon_kt}) overestimates by a factor $4$ at worse the measured dissipation. We suspect that applying the theory, which relies on isotropic turbulence, is not suited for such an oceanic application characterised by strong anisotropy. We find a better match using a rough estimate of the isotropic part of the TKE from the vertical normal Reynolds stresses only.

Finally the prediction models for the TKE and velocity are applied to estimate their wave-induced component. The spread caused by the inherent variability in the data and the model approximations are relatively high, close to the effect we wish to observe. However we believe it is still possible to exploit the results especially for stronger sea states where the impact on turbulence is more noticeable.
Follow-up work will focus on fitting relevant profiles to those wave-induced profiles, and study their dependence to the wave and flow parameters. We expect the significant wave height to play a major part, but not only. Since the injection of turbulent energy is caused by wave breaking other secondary parameters such as the flow direction, wave direction and wave period should be of importance as well.
We hope that it will contribute to a better estimation of the load induced by waves on submerged structures, as well as a better understanding of the vertical mixing in the water column, crucial in coastal ocean modelling.

\section*{Acknowledgments}{
The authors acknowledge the support of the French Agence Nationale de la Recherche (ANR), under grants ANR-21-ASM1-0003 (project MORHOC'H 2), and ANR-10-IEED-0006-07 (project HYD2M).
We also wish to acknowledge the crew of the R/V ‘Côtes de la Manche’ and the support from France Energies Marines (FEM).
}

\section*{Data availability}{
The raw and processed datasets used and analysed during the current study are available from \url{https://geoceano.intechmer.cnam.fr/programmes/anr-hyd2m/donnees-hyd2m/}, upon request of a login credential.}

\section*{Competing interests}{
No potential conflict of interest is reported by the authors.}

\section*{Contributions}{
\textbf{C Calvino}: Conceptualisation, Methodology, Software, Formal analysis, Investigation, Writing - Original Draft.
\textbf{L Furgerot, E Poizot}: Resources, Software, Investigation, Writing - Review \& Editing.
\textbf{P Bailly du Bois, A-C Bennis}: Supervision, Investigation, Writing - Review \& Editing, Project administration, Funding acquisition.}

%\appendix
%\section{List of symbols \label{symbols}}

\nomenclature{\(\rho\)}{Mean total pressure}
\nomenclature{\(\tau\)}{Mean total stress}
\nomenclature{\(P\)}{Mean total pressure}
\nomenclature{\(h\)}{Mean total depth}
\nomenclature{\(s\)}{Stream-wise coordinate}
\nomenclature{\(u, w\)}{Stream-wise and vertical mean velocity}
\nomenclature{\(u_\text{w}\)}{Stream-wise mean velocity induced by waves}
\nomenclature{\(U, W\)}{Stream-wise and vertical mean barotropic (depth-integrated) velocity}
\nomenclature{\(u', w'\)}{Stream-wise and vertical turbulent velocities}
\nomenclature{\(x, y, z\)}{Cartesian coordinates, $z$ pointing upward, origin at the sea bottom}
\nomenclature{\(u_x, u_y, u_z\)}{Cartesian mean velocities}
\nomenclature{\(u_x', u_y', u_z'\)}{Cartesian turbulent velocities}

\nomenclature{\(\nu_t\)}{Turbulent viscosity}
\nomenclature{\(l_m\)}{Prandlt's mixing length}
\nomenclature{\(\kappa\)}{von Kármán constant}
\nomenclature{\(\tau_0\)}{Mean total stress at the wall (bottom)}
\nomenclature{\(z_0\)}{Bottom roughness height}
\nomenclature{\(u_\tau\)}{Friction velocity}
\nomenclature{\(k_t\)}{Turbulent kinetic energy}
\nomenclature{\(k_{t,\text{iso}}\)}{Turbulent kinetic energy in the isotropic range}
\nomenclature{\(k_{t,\text{wave}}\)}{Turbulent kinetic energy induced by waves}
\nomenclature{\(\epsilon\)}{Turbulent dissipation}
\nomenclature{\(\mathcal{P}\)}{Turbulent production}
\nomenclature{\(c, c_\epsilon, \sigma_k\)}{One-equation turbulent viscosity closure model constants}

\nomenclature{\(u_{\tau,\text{bot}}, u_{\tau,\text{up}}\)}{Double logarithmic law bottom and upper layer friction velocities}
\nomenclature{\(z_{0,\text{bot}}, z_{0,\text{up}}\)}{Double logarithmic law bottom and upper layer roughness heights}
\nomenclature{\(z_\text{lim}\)}{Double logarithmic law bottom and upper layers separation height}

\nomenclature{\(\delta, \gamma\)}{Power law roughness and power coefficient}

\nomenclature{\(a_\tau, b_\tau\)}{Friction velocity regression slope and origin coefficients}

\nomenclature{\(\alpha, \beta\)}{TKE height regression slope and origin coefficient}
\nomenclature{\(a_\alpha, b_\alpha, p_\alpha\)}{$\alpha$ regression amplitude, origin and velocity scaling coefficients}
\nomenclature{\(a_\beta, b_\beta, p_\beta\)}{$\beta$ regression amplitude, origin and velocity scaling coefficients}
\nomenclature{\(A(z), k_{t,0}(z), p\)}{TKE regression shape, residual and velocity scaling profiles}

\nomenclature{\(\theta\)}{ADCP beam inclination}
\nomenclature{\(\psi, \phi, \xi\)}{ADCP roll, pitch and heading angles}
\nomenclature{\(k\)}{Turbulent structures wavenumber}
\nomenclature{\(b_i'\)}{Turbulent velocity along beam $i$}
\nomenclature{\(S(k), S_{ii}\)}{Turbulent velocity spectral density, total and along beam $i$}
\nomenclature{\(N(k), N_{i}\)}{Doppler noise, total and along beam $i$}
\nomenclature{\(S_w(k), S_{w,i}\)}{Wave spectral density, total and along beam $i$}
\nomenclature{\(C, C_u, C_w\)}{Vertical Kolmogorov total, stream-wise and orthogonal constants}

\nomenclature{\(m_{ij}\)}{Matrix coefficients from ENU to XYZ ADCP coordinate system}
\nomenclature{\(w_e, w_n, w_u\)}{Projection coefficients in the ENU system for the orbital velocities}
\nomenclature{\(w_x, w_y, w_z\)}{Projection coefficients in the XYZ system for the orbital velocities}
\nomenclature{\(w_{bi}\)}{Projection coefficients in $i$ beam direction for the orbital velocities}
\nomenclature{\(f_w\)}{Relative wave frequency}
\nomenclature{\(k_w\)}{Wavenumber}
\nomenclature{\(\theta_w\)}{Wave direction}

\printnomenclature

\appendix
\section{Reminders on wall turbulence theory  \label{wall_turbulence_theory}}

The law of the wall can be derived from the mean-axial-momentum equation given below, after neglecting temporal and horizontal variations, vertical advection and assuming the Boussinesq approximation:

\begin{equation}
 \dpar{\tau}{z} -  \dstr{P}{s} = 0 . \label{mean_momentum}
\end{equation}

$P$ is the mean total pressure, $s$ the stream-wise spatial coordinate, and $\tau$ the mean total shear stress defined as follows (e.g. \cite{pope2001turbulent}) after neglecting the molecular viscosity, with the over-bar indicating a time average, $u'$ is the stream-wise turbulent velocity and $w'$ the vertical turbulent velocity:
%The mean total shear stress $\tau$ is defined as follows (e. g. \cite{pope2001turbulent}) after neglecting the molecular viscosity: 

\begin{equation}
 \tau = - \rho \overline{u'w'}.
\end{equation}

%\begin{equation}
% \tau = - \rho \overline{u'w'} ,\qquad u^t = u + u' ,\qquad w^t = w + w'.
%\end{equation}

%In the above $u'$ is the stream-wise turbulent velocity and $w'$ the vertical turbulent velocity, obtained from the Reynolds decomposition of the total velocities $(u^t, w^t)$ in mean $(u,w)$ and turbulent velocities.
%If required for clarity the velocities are also written with a subscript convention, for instance the mean Cartesian velocities are $(u_x,u_y,u_z)$ in the $(x,y,z)$ directions with $z$ pointing upwards and equal to $0$ at the sea bottom.
%The depth-integrated velocities are indicated with an upper-case letter.

The shear stress is then modelled with a turbulent viscosity model as follows, where $\nu_t$ is the turbulent viscosity, $\kappa$ the von Kármán constant, $l_m$ the Prandlt's mixing length, $u_\tau$ the friction velocity, $z_0$ the bottom roughness height and $\tau_0$ the shear stress at the bottom: 

\begin{align}
& \tau = \rho \nu_t \dpar{u}{z} , \label{tau}\\
& \nu_t = l_m u_\tau , \label{nu}\\
& l_m = \kappa (z - z_0) , \label{lm}\\
& u_\tau = \sqrt{\tau_0 / \rho}. \label{ut}
\end{align}

From Eq. (\ref{mean_momentum}) we obtain that the shear stress is a linear function of $z$. In practice it is found constant near the bottom and equal to its wall value $\tau_0$ within a bottom layer near the seabed, and from Eqs. (\ref{tau}--\ref{ut}) the law of the wall is found:

\begin{equation}
 u = \frac{u_\tau}{\kappa} \ln{\left(\frac{z}{z_0}\right)} . \label{log}
\end{equation}

Upper in the water column the shear stress is expected to decrease with depth, often assumed linearly. The log law is then not supposed to hold but it is often found to give a good approximation. 
%Several corrections appearing in the upper layer are mentioned in Section \ref{section_method}, with the results discussed in Section \ref{section_profiles}.

The friction velocity scaling proposed in Eq. (\ref{ut}) with the shear stress $\tau$ is an arbitrary choice, often made when only the velocity profile is of interest. Without any incidence on the results it can be better scaled with the turbulent kinetic energy $k_t$ as follows, introducing a constant $c$ determined later:

\begin{equation}
 u_\tau = c k_t^{1/2} , \qquad \text{with} \qquad k_t = \frac{1}{2} \left( \overline{u_x'u_x'} + \overline{u_y'u_y'} + \overline{u_z'u_z'} \right) . \label{ut_kt}
\end{equation}

The logarithmic law region is characterised by a constant shear stress, or equivalently by a constant TKE, referring to the above relation. It yields the classic relationship $\overline{u'w'} = c^2 k_t$ valid within the logarithmic layer (e.g. \cite{pope2001turbulent}).

The scaling with the TKE is better suited for a one-equation turbulent viscosity closure model \cite{rascle2006drift,feddersen2005effect}, for instance:

\begin{equation}
\dpar{k_t}{t} + u_z \dpar{k_t}{z} = \dpar{}{z} \left(\frac{\nu_t}{\sigma_k} \dpar{k_t}{z} \right) + \nu_t \left( \left(\dpar{u_x}{z}\right)^2 + \left(\dpar{u_y}{z}\right)^2 \right) - \epsilon . \label{tke_balance}
\end{equation}

With:

\begin{align}
 & \epsilon = c_\epsilon k_t^{3/2}/l_m , \label{epsilon_kt}\\
 & \nu_t = c k_t^{1/2} l_m . \label{nu_kt}
\end{align}

This balance equation for the TKE is already a simplified version where turbulent statistics are notably assumed uniform in the horizontal directions. The horizontal advection and horizontal TKE diffusion are neglected and only the vertical mean momentum shear is producing TKE. The terms on the right-hand-side of Eq. (\ref{tke_balance}) are from left to right the transport of TKE modelled by a gradient-diffusion mechanism, the shear production and the turbulent dissipation.
Since we do not provide an equation for the dissipation we evaluate it through the TKE itself with Eq. (\ref{epsilon_kt}). The mixing length $l_m$ is still unknown and it is provided with Eq. (\ref{lm}).
Referring to \cite{rascle2006drift} and \cite{pope2001turbulent}, the constants take the following values, they are initially found through validation against laboratory channel experiments:

\begin{equation}
 c_\epsilon = c^3 = 0.17 ,\qquad c = 0.55 ,\qquad \sigma_k = 1.95 .
\end{equation}

\section{Computation of the wave orbital spectra \label{wave_orbital_app}}

The linear theory is used to evaluate the wave orbital spectra, but we need to project the spectra accordingly in the beam directions keeping in mind that only the energy levels are relatable when conducting spatial projections, and not the velocities \cite{dewey2007reynolds}.

We start by defining the coefficient for the eastward, northward and upward orbital velocities $(w_e, w_n, w_u)$ as follows, with $f_w$ the relative wave frequency, $k_w$ is the wavenumber for the given wave frequency and wave direction $\theta_w$ (in Cartesian convention), $H$ the mean total water depth, and $z$ is still the height above seabed:

\begin{align}
& w_e = \frac{2\pi f_w  \cosh{(k_w z)}}{\sinh{(k_w h)}} \cos{\theta_w} , \\
& w_n = \frac{2\pi f_w  \cosh{(k_w z)}}{\sinh{(k_w h)}} \sin{\theta_w} , \\
& w_u = \frac{2\pi f_w  \sinh{(k_w z)}}{\sinh{(k_w h)}} .
\end{align}

Those coefficients squared can be used in conjunction with the wave spectra to infer the wave orbital spectra in the eastward, northward and upward directions. However we wish to obtain the wave orbital spectra in the beam directions so further projection coefficients are needed. Below are reminded the matrix coefficients $m_{ij}$ from the East-North-Up coordinate system to the ADCP local XYZ coordinate system, with $(\psi, \phi, \xi)$ the roll, pitch and heading angles respectively:

\begin{align}
& m_{11} = \cos{\xi}\cos{\psi} - \sin{\xi}\sin{\phi}\sin{\psi} , \\
& m_{12} = -\sin{\xi}\cos{\phi} , \\
& m_{13} = -\cos{\xi}\sin{\psi} - \sin{\xi}\sin{\phi}\cos{\psi} , \\
& m_{21} = \sin{\xi}\cos{\psi} + \cos{\xi}\sin{\phi}\sin{\psi} , \\
& m_{22} = \cos{\xi}\cos{\phi} , \\
& m_{23} = -\sin{\xi}\sin{\psi} + \cos{\xi}\sin{\phi}\cos{\psi} , \\
& m_{31} = \cos{\phi}\sin{\psi} , \\
& m_{32} = -\sin{\phi} , \\
& m_{33} = \cos{\phi}\cos{\psi} .
\end{align}

The spectra coefficients $(w_x, w_y, w_z)$ for the projection in the XYZ coordinate system are then as follows:

\begin{align}
& w_x = m_{11}^2 w_e^2 + m_{12}^2 w_n^2 + m_{13}^2 w_u^2 + 2 m_{11} m_{12} w_e w_n ,\\
& w_y = m_{21}^2 w_e^2 + m_{22}^2 w_n^2 + m_{23}^2 w_u^2 + 2 m_{21} m_{22} w_e w_n ,\\
& w_z = m_{31}^2 w_e^2 + m_{32}^2 w_n^2 + m_{33}^2 w_u^2 + 2 m_{31} m_{32} w_e w_n .
\end{align}

And finally the spectra coefficients $(w_{b1}, w_{b2}, w_{b3}, w_{b4}, w_{b5})$ for the projection in the four inclined beam directions and vertical direction are as follows, with $\theta$ the beam inclination:

\begin{align}
& w_{b1} = w_x \sin\theta^2 + w_z \cos\theta^2 \nonumber\\
& + \left( m_{11}m_{31} w_e^2 + m_{12}m_{32} w_n^2 + m_{13}m_{33} w_u^2 + (m_{11}m_{32}+m_{12}m_{31}) w_e w_n \right) \frac{\sin{2\theta}}{2} ,\\
& w_{b2} = w_x \sin\theta^2 + w_z \cos\theta^2 \nonumber\\
& - \left( m_{11}m_{31} w_e^2 + m_{12}m_{32} w_n^2 + m_{13}m_{33} w_u^2 + (m_{11}m_{32}+m_{12}m_{31}) w_e w_n \right) \frac{\sin{2\theta}}{2} ,\\
& w_{b3} = w_y \sin\theta^2 + w_z \cos\theta^2 \nonumber\\
& - \left( m_{21}m_{31} w_e^2 + m_{22}m_{32} w_n^2 + m_{23}m_{33} w_u^2 + (m_{21}m_{32}+m_{22}m_{31}) w_e w_n \right) \frac{\sin{2\theta}}{2} ,\\
& w_{b4} = w_y \sin\theta^2 + w_z \cos\theta^2 \nonumber\\
& + \left( m_{21}m_{31} w_e^2 + m_{22}m_{32} w_n^2 + m_{23}m_{33} w_u^2 + (m_{21}m_{32}+m_{22}m_{31}) w_e w_n \right) \frac{\sin{2\theta}}{2} ,\\
& w_{b5} = w_z .
\end{align}

The wave orbital spectra projected in each five beam directions $S_{w,i}$ are then computed as follows with $E(f_w, \theta_w)$ the wave energy spectral density, averaging over the wave direction:

\begin{align}
& S_{w,i}(f_w) = \int_0^{2 \pi} w_{b,i} (f_w, \theta_w) E(f_w, \theta_w) \,\mathrm{d}\theta_w.
\end{align}

\section{Definition of the used statistics \label{statistics_def}}

The statistics used in this paper are defined below, they correspond to depth-average values for each individual profile, they are then averaged over profiles with possible constraints on the flow characteristics. $X_{\text{obs},i}$ corresponds to the measured value at bin number $i$ while $X_{\text{mod},i}$ corresponds to the estimate or prediction, and $D$ is the number of bins where data is available:

\begin{align}
& \text{Bias} = \frac{1}{D} \sum_{i=1}^{D} \left(X_{\text{mod},i} - X_{\text{obs},i}\right) .\\
& \text{RMSE} = \sqrt{ \frac{1}{D} \sum_{i=1}^{D} \left(X_{\text{mod},i} - X_{\text{obs},i}\right)^2} .\\*
& \text{NRMSE} = \sqrt{ \frac{1}{D} \sum_{i=1}^{D} \left(X_{\text{mod},i} - X_{\text{obs},i}\right)^2} \Bigg/ \frac{1}{D} \sum_{i=1}^{D} X_{\text{obs},i} .\\
& R = \frac{\sum_{i=1}^{D} \left(X_{\text{mod},i} - \frac{1}{D} \sum_{i=1}^{D} X_{\text{mod},i}\right) \left(X_{\text{obs},i} - \frac{1}{D} \sum_{i=1}^{D} X_{\text{obs},i}\right)}{\sqrt{\sum_{i=1}^{D} \left(X_{\text{mod},i} - \frac{1}{D} \sum_{i=1}^{D} X_{\text{mod},i}\right)^2  \sum_{i=1}^{D} \left(X_{\text{obs},i} - \frac{1}{D} \sum_{i=1}^{D} X_{\text{obs},i}\right)^2}}.
\end{align}

\newpage
\bibliographystyle{elsarticle-num} 
\bibliography{Bib.bib}

\begin{thebibliography}{10}
\expandafter\ifx\csname url\endcsname\relax
  \def\url#1{\texttt{#1}}\fi
\expandafter\ifx\csname urlprefix\endcsname\relax\def\urlprefix{URL }\fi
\expandafter\ifx\csname href\endcsname\relax
  \def\href#1#2{#2} \def\path#1{#1}\fi

\bibitem{mercier2020numerical}
P.~Mercier, M.~Grondeau, S.~Guillou, J.~Thiébot, E.~Poizot, Numerical study of
  the turbulent eddies generated by the seabed roughness. case study at a tidal
  power site, Applied Ocean Research 97 (2020) 102082.
\newblock \href {https://doi.org/10.1016/j.apor.2020.102082}
  {\path{doi:10.1016/j.apor.2020.102082}}.

\bibitem{neill2021review}
S.~P. Neill, K.~A. Haas, J.~Thiébot, Z.~Yang, A review of tidal
  energy-resource, feedbacks, and environmental interactions, Journal of
  Renewable and Sustainable Energy 13~(6) (2021) 062702.
\newblock \href {https://doi.org/10.1063/5.0069452}
  {\path{doi:10.1063/5.0069452}}.

\bibitem{togneri2016micrositing}
M.~Togneri, I.~Masters, Micrositing variability and mean flow scaling for
  marine turbulence in {R}amsey {S}ound, Journal of Ocean Engineering and
  Marine Energy 2~(1) (2016) 35--46.
\newblock \href {https://doi.org/10.1007/s40722-015-0036-0}
  {\path{doi:10.1007/s40722-015-0036-0}}.

\bibitem{thomson2012measurements}
J.~Thomson, B.~Polagye, V.~Durgesh, M.~C. Richmond, Measurements of turbulence
  at two tidal energy sites in {P}uget {S}ound, {WA}, IEEE Journal of Oceanic
  Engineering 37~(3) (2012) 363--374.
\newblock \href {https://doi.org/10.1109/JOE.2012.2191656}
  {\path{doi:10.1109/JOE.2012.2191656}}.

\bibitem{baillydubois2020alderney}
P.~{Bailly du Bois}, F.~Dumas, M.~Morillon, L.~Furgerot, C.~Voiseux, E.~Poizot,
  Y.~Méar, A.-C. Bennis, The {A}lderney {R}ace: general hydrodynamic and
  particular features, Philosophical Transactions of the Royal Society A:
  Mathematical, Physical and Engineering Sciences 378~(2178) (2020) 20190492.
\newblock \href {https://doi.org/10.1098/rsta.2019.0492}
  {\path{doi:10.1098/rsta.2019.0492}}.

\bibitem{bennis2020numerical}
A.-C. Bennis, L.~Furgerot, P.~{Bailly du Bois}, F.~Dumas, T.~Odaka,
  C.~Lathuilière, J.-F. Filipot, Numerical modelling of three-dimensional
  wave-current interactions in complex environment: Application to {A}lderney
  {R}ace, Applied Ocean Research 95 (2020) 102021.
\newblock \href {https://doi.org/10.1016/j.apor.2019.102021}
  {\path{doi:10.1016/j.apor.2019.102021}}.

\bibitem{pope2001turbulent}
S.~B. Pope, Turbulent flows, IOP Publishing, 2001.

\bibitem{perlin2005lawofthewall}
A.~Perlin, J.~N. Moum, J.~M. Klymak, M.~D. Levine, T.~Boyd, P.~M. Kosro, A
  modified law-of-the-wall applied to oceanic bottom boundary layers, Journal
  of Geophysical Research: Oceans 110~(C10) (2005).
\newblock \href {https://doi.org/10.1029/2004JC002310}
  {\path{doi:10.1029/2004JC002310}}.

\bibitem{lueck1997logarithmic}
R.~G. Lueck, Y.~Lu, The logarithmic layer in a tidal channel, Continental Shelf
  Research 17~(14) (1997) 1785--1801.
\newblock \href {https://doi.org/10.1016/S0278-4343(97)00049-6}
  {\path{doi:10.1016/S0278-4343(97)00049-6}}.

\bibitem{nezu1993turbulence}
I.~Nezu, H.~Nakagawa, Turbulence in open-channel flows, Taylor \& Francis,
  1993.
\newblock \href {https://doi.org/10.1201/9780203734902}
  {\path{doi:10.1201/9780203734902}}.

\bibitem{lewis2017characteristics}
M.~Lewis, S.~Neill, P.~Robins, M.~Hashemi, S.~Ward, Characteristics of the
  velocity profile at tidal-stream energy sites, Renewable Energy 114 (2017)
  258--272, wave and Tidal Resource Characterization.
\newblock \href {https://doi.org/10.1016/j.renene.2017.03.096}
  {\path{doi:10.1016/j.renene.2017.03.096}}.

\bibitem{thiebaut2017asymmetry}
M.~Thiébaut, A.~Sentchev, Asymmetry of tidal currents off the {W}.{B}rittany
  coast and assessment of tidal energy resource around the {U}shant {I}sland,
  Renewable Energy 105 (2017) 735--747.
\newblock \href {https://doi.org/10.1016/j.renene.2016.12.082}
  {\path{doi:10.1016/j.renene.2016.12.082}}.

\bibitem{togneri2017comparison}
M.~Togneri, M.~Lewis, S.~Neill, I.~Masters, Comparison of {ADCP} observations
  and {3D} model simulations of turbulence at a tidal energy site, Renewable
  Energy 114 (2017) 273--282, wave and Tidal Resource Characterization.
\newblock \href {https://doi.org/10.1016/j.renene.2017.03.061}
  {\path{doi:10.1016/j.renene.2017.03.061}}.

\bibitem{thiebaut2020assessing}
M.~Thiébaut, J.-F. Filipot, C.~Maisondieu, G.~Damblans, R.~Duarte, E.~Droniou,
  S.~Guillou, Assessing the turbulent kinetic energy budget in an energetic
  tidal flow from measurements of coupled {ADCPs}, Philosophical Transactions
  of the Royal Society A: Mathematical, Physical and Engineering Sciences
  378~(2178) (2020) 20190496.
\newblock \href {https://doi.org/10.1098/rsta.2019.0496}
  {\path{doi:10.1098/rsta.2019.0496}}.

\bibitem{furgerot2020measurements}
L.~Furgerot, A.~Sentchev, P.~{Bailly du Bois}, G.~Lopez, M.~Morillon,
  E.~Poizot, Y.~Méar, A.-C. Bennis, One year of measurements in {A}lderney
  {R}ace: preliminary results from database analysis, Philosophical
  Transactions of the Royal Society A: Mathematical, Physical and Engineering
  Sciences 378~(2178) (2020) 20190625.
\newblock \href {https://doi.org/10.1098/rsta.2019.0625}
  {\path{doi:10.1098/rsta.2019.0625}}.

\bibitem{mcmillan2016rates}
J.~M. McMillan, A.~E. Hay, R.~G. Lueck, F.~Wolk, Rates of dissipation of
  turbulent kinetic energy in a high {R}eynolds number tidal channel, Journal
  of Atmospheric and Oceanic Technology 33~(4) (2016) 817 -- 837.
\newblock \href {https://doi.org/10.1175/JTECH-D-15-0167.1}
  {\path{doi:10.1175/JTECH-D-15-0167.1}}.

\bibitem{guerra2017turbulence}
M.~Guerra, J.~Thomson, {T}urbulence measurements from five-beam {A}coustic
  {D}oppler {C}urrent {P}rofilers, Journal of Atmospheric and Oceanic
  Technology 34~(6) (2017) 1267 -- 1284.
\newblock \href {https://doi.org/10.1175/JTECH-D-16-0148.1}
  {\path{doi:10.1175/JTECH-D-16-0148.1}}.

\bibitem{mcmillan2017spectral}
J.~M. McMillan, A.~E. Hay, Spectral and structure function estimates of
  turbulence dissipation rates in a high-flow tidal channel using broadband
  {ADCPs}, Journal of Atmospheric and Oceanic Technology 34~(1) (2017) 5 -- 20.
\newblock \href {https://doi.org/10.1175/JTECH-D-16-0131.1}
  {\path{doi:10.1175/JTECH-D-16-0131.1}}.

\bibitem{krug2017revisiting}
D.~Krug, J.~Philip, I.~Marusic, Revisiting the law of the wake in wall
  turbulence, Journal of Fluid Mechanics 811 (2017) 421–435.
\newblock \href {https://doi.org/10.1017/jfm.2016.788}
  {\path{doi:10.1017/jfm.2016.788}}.

\bibitem{sanford1999turbulent}
T.~B. Sanford, R.-C. Lien, Turbulent properties in a homogeneous tidal bottom
  boundary layer, Journal of Geophysical Research: Oceans 104~(C1) (1999)
  1245--1257.
\newblock \href {https://doi.org/10.1029/1998JC900068}
  {\path{doi:10.1029/1998JC900068}}.

\bibitem{trowbridge1999nearbottom}
J.~H. Trowbridge, W.~R. Geyer, M.~M. Bowen, A.~J. Williams, Near-bottom
  turbulence measurements in a partially mixed estuary: Turbulent energy
  balance, velocity structure, and along-channel momentum balance, Journal of
  Physical Oceanography 29~(12) (1999) 3056 -- 3072.
\newblock \href
  {https://doi.org/10.1175/1520-0485(1999)029<3056:NBTMIA>2.0.CO;2}
  {\path{doi:10.1175/1520-0485(1999)029<3056:NBTMIA>2.0.CO;2}}.

\bibitem{trowbridge2018boundary}
J.~H. Trowbridge, S.~J. Lentz, The bottom boundary layer, Annual Review of
  Marine Science 10~(1) (2018) 397--420.
\newblock \href {https://doi.org/10.1146/annurev-marine-121916-063351}
  {\path{doi:10.1146/annurev-marine-121916-063351}}.

\bibitem{deserio2014streamwise}
F.~De~Serio, M.~Mossa, Streamwise velocity profiles in coastal currents,
  Environmental Fluid Mechanics 14 (2014) 895--918.
\newblock \href {https://doi.org/10.1007/s10652-014-9338-3}
  {\path{doi:10.1007/s10652-014-9338-3}}.

\bibitem{jimenez2004turbulent}
J.~Jim\'{e}nez, Turbulent flows over rough walls, Annual Review of Fluid
  Mechanics 36~(1) (2004) 173--196.
\newblock \href {https://doi.org/10.1146/annurev.fluid.36.050802.122103}
  {\path{doi:10.1146/annurev.fluid.36.050802.122103}}.

\bibitem{allen2007turbulent}
J.~Allen, M.~Shockling, G.~Kunkel, A.~Smits, Turbulent flow in smooth and rough
  pipes, Philosophical Transactions of the Royal Society A: Mathematical,
  Physical and Engineering Sciences 365~(1852) (2007) 699--714.
\newblock \href {https://doi.org/10.1098/rsta.2006.1939}
  {\path{doi:10.1098/rsta.2006.1939}}.

\bibitem{goring2002despiking}
D.~G. Goring, V.~I. Nikora, Despiking {A}coustic {D}oppler {V}elocimeter data,
  Journal of Hydraulic Engineering 128~(1) (2002) 117--126.
\newblock \href {https://doi.org/10.1061/(ASCE)0733-9429(2002)128:1(117)}
  {\path{doi:10.1061/(ASCE)0733-9429(2002)128:1(117)}}.

\bibitem{dewey2007reynolds}
R.~Dewey, S.~Stringer, Reynolds stresses and turbulent kinetic energy estimates
  from various {ADCP} beam configurations: {Theory}, unpublished manuscript
  (2015).
\newblock \href {https://doi.org/10.13140/RG.2.1.1042.8002}
  {\path{doi:10.13140/RG.2.1.1042.8002}}.

\bibitem{thiebaut2020comprehensive}
M.~Thi{é}baut, J.-F. Filipot, C.~Maisondieu, G.~Damblans, R.~Duarte,
  E.~Droniou, N.~Chaplain, S.~Guillou, A comprehensive assessment of turbulence
  at a tidal-stream energy site influenced by wind-generated ocean waves,
  Energy 191 (2020) 116550.
\newblock \href {https://doi.org/10.1016/j.energy.2019.116550}
  {\path{doi:10.1016/j.energy.2019.116550}}.

\bibitem{durgesh2014noise}
V.~Durgesh, J.~Thomson, M.~C. Richmond, B.~L. Polagye, Noise correction of
  turbulent spectra obtained from {A}coustic {D}oppler {V}elocimeters, Flow
  Measurement and Instrumentation 37 (2014) 29--41.
\newblock \href {https://doi.org/10.1016/j.flowmeasinst.2014.03.001}
  {\path{doi:10.1016/j.flowmeasinst.2014.03.001}}.

\bibitem{filipot2015wave}
J.-F. Filipot, M.~Prevosto, C.~Maisondieu, M.~Le~Boulluec, J.~Thomson, Wave and
  turbulence measurements at a tidal energy site, in: 2015 IEEE/OES Eleveth
  Current, Waves and Turbulence Measurement (CWTM), 2015, pp. 1--9.
\newblock \href {https://doi.org/10.1109/CWTM.2015.7098128}
  {\path{doi:10.1109/CWTM.2015.7098128}}.

\bibitem{terray1999measuring}
E.~A. Terray, B.~H. Brumley, B.~Strong, Measuring waves and currents with an
  upward-looking {ADCP}, Proceedings of the IEEE Sixth Working Conference on
  Current Measurement (1999) 66--71\href
  {https://doi.org/10.1109/CCM.1999.755216}
  {\path{doi:10.1109/CCM.1999.755216}}.

\bibitem{furgerot2019highresolution}
L.~Furgerot, Y.~Poprawski, M.~Violet, E.~Poizot, P.~{Bailly du Bois},
  M.~Morillon, Y.~Méar, High-resolution bathymetry of the {A}lderney {R}ace
  and its geological and sedimentological description ({R}az {B}lanchard,
  northwest {F}rance), Journal of Maps 15~(2) (2019) 708--718.
\newblock \href {https://doi.org/10.1080/17445647.2019.1657510}
  {\path{doi:10.1080/17445647.2019.1657510}}.

\bibitem{stacey1999measurements}
M.~T. Stacey, S.~G. Monismith, J.~R. Burau, Measurements of {R}eynolds stress
  profiles in unstratified tidal flow, Journal of Geophysical Research: Oceans
  104~(C5) (1999) 10933--10949.
\newblock \href {https://doi.org/10.1029/1998JC900095}
  {\path{doi:10.1029/1998JC900095}}.

\bibitem{cheng1999estimates}
R.~T. Cheng, C.-H. Ling, J.~W. Gartner, P.~F. Wang, Estimates of bottom
  roughness length and bottom shear stress in {S}outh {S}an {F}rancisco {B}ay,
  {C}alifornia, Journal of Geophysical Research: Oceans 104~(C4) (1999)
  7715--7728.
\newblock \href {https://doi.org/doi.org/10.1029/1998JC900126}
  {\path{doi:doi.org/10.1029/1998JC900126}}.

\bibitem{bauer1992sources}
B.~O. Bauer, D.~J. Sherman, J.~F. Wolcott, Sources of uncertainty in shear
  stress and roughness length estimates derived from velocity profiles, The
  Professional Geographer 44~(4) (1992) 453--464.
\newblock \href {https://doi.org/10.1111/j.0033-0124.1992.00453.x}
  {\path{doi:10.1111/j.0033-0124.1992.00453.x}}.

\bibitem{mercier2021impact}
P.~Mercier, S.~Guillou, The impact of the seabed morphology on turbulence
  generation in a strong tidal stream, Physics of Fluids 33~(5) (2021) 055125.
\newblock \href {https://doi.org/10.1063/5.0047791}
  {\path{doi:10.1063/5.0047791}}.

\bibitem{mercier2021turbulence}
P.~Mercier, M.~Thiébaut, S.~Guillou, C.~Maisondieu, E.~Poizot, A.~Pieterse,
  J.~Thiébot, J.-F. Filipot, M.~Grondeau, Turbulence measurements: An
  assessment of {A}coustic {D}oppler {C}urrent {P}rofiler accuracy in rough
  environment, Ocean Engineering 226 (2021) 108819.
\newblock \href {https://doi.org/10.1016/j.oceaneng.2021.108819}
  {\path{doi:10.1016/j.oceaneng.2021.108819}}.

\bibitem{feddersen2007direct}
F.~Feddersen, A.~J. Williams, Direct estimation of the {R}eynolds stress
  vertical structure in the nearshore, Journal of Atmospheric and Oceanic
  Technology 24~(1) (2007) 102 -- 116.
\newblock \href {https://doi.org/10.1175/JTECH1953.1}
  {\path{doi:10.1175/JTECH1953.1}}.

\bibitem{bourgoin2020turbulence}
A.~C.~L. Bourgoin, S.~S. Guillou, J.~Thiébot, R.~Ata, Turbulence
  characterization at a tidal energy site using large-eddy simulations: case of
  the {A}lderney {R}ace, Philosophical Transactions of the Royal Society A:
  Mathematical, Physical and Engineering Sciences 378~(2178) (2020) 20190499.
\newblock \href {https://doi.org/10.1098/rsta.2019.0499}
  {\path{doi:10.1098/rsta.2019.0499}}.

\bibitem{fugerot2018velocity}
L.~Furgerot, P.~B. du~Bois, Y.~Méar, M.~Morillon, E.~Poizot, A.-C. Bennis,
  Velocity profile variability at a tidal-stream energy site ({A}ldemey {R}ace,
  {F}rance): From short (second) to yearly time scales, in: 2018 OCEANS -
  MTS/IEEE Kobe Techno-Oceans (OTO), 2018, pp. 1--8.
\newblock \href {https://doi.org/10.1109/OCEANSKOBE.2018.8559326}
  {\path{doi:10.1109/OCEANSKOBE.2018.8559326}}.

\bibitem{krogstad2005experimental}
P.-{\AA}. Krogstad, H.~I. Andersson, O.~M. Bakken, A.~Ashrafian, An
  experimental and numerical study of channel flow with rough walls, Journal of
  Fluid Mechanics 530 (2005) 327–352.
\newblock \href {https://doi.org/10.1017/S0022112005003824}
  {\path{doi:10.1017/S0022112005003824}}.

\bibitem{chatzikyriakou2015turbulent}
D.~Chatzikyriakou, J.~Buongiorno, D.~Caviezel, D.~Lakehal, {DNS} and {LES} of
  turbulent flow in a closed channel featuring a pattern of hemispherical
  roughness elements, International Journal of Heat and Fluid Flow 53 (2015)
  29--43.
\newblock \href {https://doi.org/10.1016/j.ijheatfluidflow.2015.01.002}
  {\path{doi:10.1016/j.ijheatfluidflow.2015.01.002}}.

\bibitem{milne2013characteristics}
I.~A. Milne, R.~N. Sharma, R.~G.~J. Flay, S.~Bickerton, Characteristics of the
  turbulence in the flow at a tidal stream power site, Philosophical
  Transactions of the Royal Society A: Mathematical, Physical and Engineering
  Sciences 371~(1985) (2013) 20120196.
\newblock \href {https://doi.org/10.1098/rsta.2012.0196}
  {\path{doi:10.1098/rsta.2012.0196}}.

\bibitem{groeneweg1998changes}
J.~Groeneweg, G.~Klopman, Changes of the mean velocity profiles in the combined
  wave–current motion described in a {GLM} formulation, Journal of Fluid
  Mechanics 370 (1998) 271–296.
\newblock \href {https://doi.org/10.1017/S0022112098002018}
  {\path{doi:10.1017/S0022112098002018}}.

\bibitem{olabarrieta2010effects}
M.~Olabarrieta, R.~Medina, S.~Castanedo, Effects of wave–current interaction
  on the current profile, Coastal Engineering 57~(7) (2010) 643--655.
\newblock \href {https://doi.org/10.1016/j.coastaleng.2010.02.003}
  {\path{doi:10.1016/j.coastaleng.2010.02.003}}.

\bibitem{rascle2006drift}
N.~Rascle, F.~Ardhuin, E.~A. Terray, Drift and mixing under the ocean surface:
  A coherent one-dimensional description with application to unstratified
  conditions, Journal of Geophysical Research: Oceans 111~(C3) (2006).
\newblock \href {https://doi.org/10.1029/2005JC003004}
  {\path{doi:10.1029/2005JC003004}}.

\bibitem{feddersen2005effect}
F.~Feddersen, J.~Trowbridge, The effect of wave breaking on surf-zone
  turbulence and alongshore currents: A modeling study, Journal of Physical
  Oceanography 35~(11) (2005) 2187--2203.
\newblock \href {https://doi.org/10.1175/JPO2800.1}
  {\path{doi:10.1175/JPO2800.1}}.

\end{thebibliography}

\end{document}